\pdfoutput=1
\documentclass{article}
\usepackage{soul}
\usepackage{lineno,hyperref}
\usepackage{multirow}
\usepackage[final,nonatbib]{nips_2016}
\usepackage[numbers]{natbib}

\usepackage[utf8]{inputenc} 
\usepackage[T1]{fontenc}    
\usepackage{hyperref}       
\usepackage{url}            
\usepackage{booktabs}             
\usepackage{nicefrac}       
\usepackage{microtype}      
\usepackage[pdftex]{graphicx}

\usepackage{mathptmx}
\usepackage[cmex10]{amsmath}
\usepackage{amsfonts}

\usepackage{array}
\newcolumntype{L}[1]{>{\raggedright\let\newline\\\arraybackslash\hspace{0pt}}m{#1}}
\newcolumntype{C}[1]{>{\centering\let\newline\\\arraybackslash\hspace{0pt}}m{#1}}
\newcolumntype{R}[1]{>{\raggedleft\let\newline\\\arraybackslash\hspace{0pt}}m{#1}}

\title{Measuring and correcting wobble in large-scale transmission radiography}

\author{
  Thomas W.~Rogers$^{1,3}$ \quad James~Ollier$^{2}$ \quad Edward J.~Morton$^{2}$  \quad Lewis D.~Griffin$^{1*}$\\\\
  $^1$Department of Computer Science, University College London, London, UK\\
  $^2$Rapiscan Systems Ltd., Stroke-on-Trent, UK\\
  $^3$Department of Security and Crime Sciences, University College London, London, UK\\\\
  \small{$^{*}$Correspondence: l.griffin@cs.ucl.ac.uk}
}

\begin{document}

\maketitle

\begin{abstract}
\noindent \textbf{BACKGROUND:}
Large-scale transmission radiography scanners are used to image vehicles and cargo containers. Acquired images are inspected for threats by a human operator or a computer algorithm. To make accurate detections, it is important that image values are precise. However, due to the scale of such systems, they can be mechanically unstable, causing the imaging array to wobble during a scan. This leads to an effective loss of precision in the captured image. \\
\textbf{OBJECTIVE:}
We consider the measurement of wobble and amelioration of the consequent loss of image precision. \\
\textbf{METHODS:}
Following our previous work, we use Beam Position Detectors (BPDs) to measure the cross-sectional profile of the X-ray beam, allowing for estimation, and thus correction of wobble. We propose: (i) a model of image formation with a wobbling detector array; (ii) a method of wobble correction derived from this model; (iii) methods for calibrating sensor sensitivities and relative offsets; (iv) a Random Regression Forest based method for instantaneous estimation of detector wobble, and (v) using these estimates to apply corrections to captured images of difficult scenes. \\
\textbf{RESULTS:}
We show that these methods are able to correct for $87\%$ of image error due wobble, and when applied to difficult images, a significant visible improvement in the intensity-windowed image quality is observed.\\
\textbf{CONCLUSIONS:}
The method improves the precision of wobble affected images, which should help improve detection of threats and the identification of different materials in the image.
\end{abstract}

\section{Introduction}
%\mainmatter
Large-scale transmission radiography has become an essential tool for detecting threats inside vehicles and cargo containers. Threats may be related to customs (drugs, counterfeit goods, banned imports, stowaways, stolen cars) or security (firearms, improvised explosive devices, special nuclear materials, missiles)~\cite{Vogel2007e,Vogel2007f,Vogel2007h,Vogel2007a,Jaccard2014,Speller2001}. Transmission radiography systems have become a mainstay of customs and border agencies around the world, and are finding increasing use in areas such as defence and the security of critical infrastructure, ports and events.  

\begin{figure}[!h]
   \centering
\includegraphics[width=0.9\textwidth]{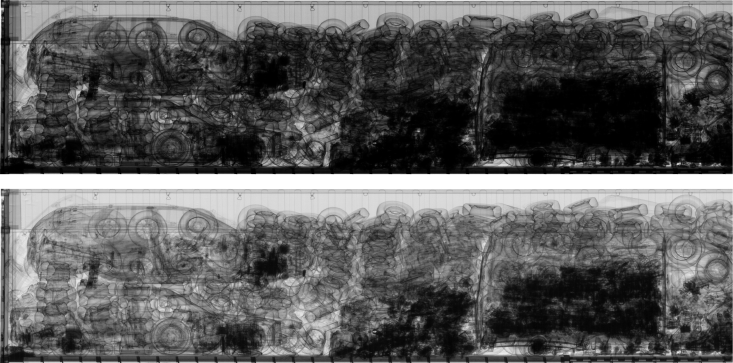}
     \caption{A raw transmission X-ray image of a cargo container containing vehicles and vehicle parts (top) and an intensity manipulated version (bottom). Intensity manipulation is often used to reveal details in the image when searching for threats.}
  \label{fig:cargoImage}
\end{figure}

Images acquired by large-scale transmission radiography (Fig.~\ref{fig:cargoImage}) are inspected by a human operator or increasingly by computer algorithm~\cite{Jaccard2014,Rogers2015}. Detection of threats by operators is assisted by intensity manipulation (windowing, logarithms, histogram equalisation) and pseudo-colouring~\cite{Chen2005a}. Additionally, scanners that acquire images at multiple photon energies permit material separation~\cite{Ogorodnikov2002} to be visualised based on differential absorption. On the basis of this visual inspection, the operator will either flag the vehicle for manual inspection or allow it to continue unimpeded. 

In order to detect threats, high spatial resolution and accurate image values are required~\cite{reed2007x} The former, because threats may be small, and the latter because threats may be shielded by other cargo or only revealed by subtle differential absorption. State-of-the-art {transmission} systems offer imaging of vehicle contents at resolutions of a few mm/pixel~\cite{Liu2008} and precisions of 16 bits. In some systems, mechanical instability (wobble) leads to effective loss of precision. {Whilst large-scale X-ray Computed Tomography (CT) could alleviate the issues of wobble and shielding, such systems are not widely deployed because they are too expensive and inefficient to be competitive~\cite{calvert2013preliminary,zentai2010x}.}

\begin{figure}[htbp]
   \centering
\includegraphics[width=0.9\textwidth]{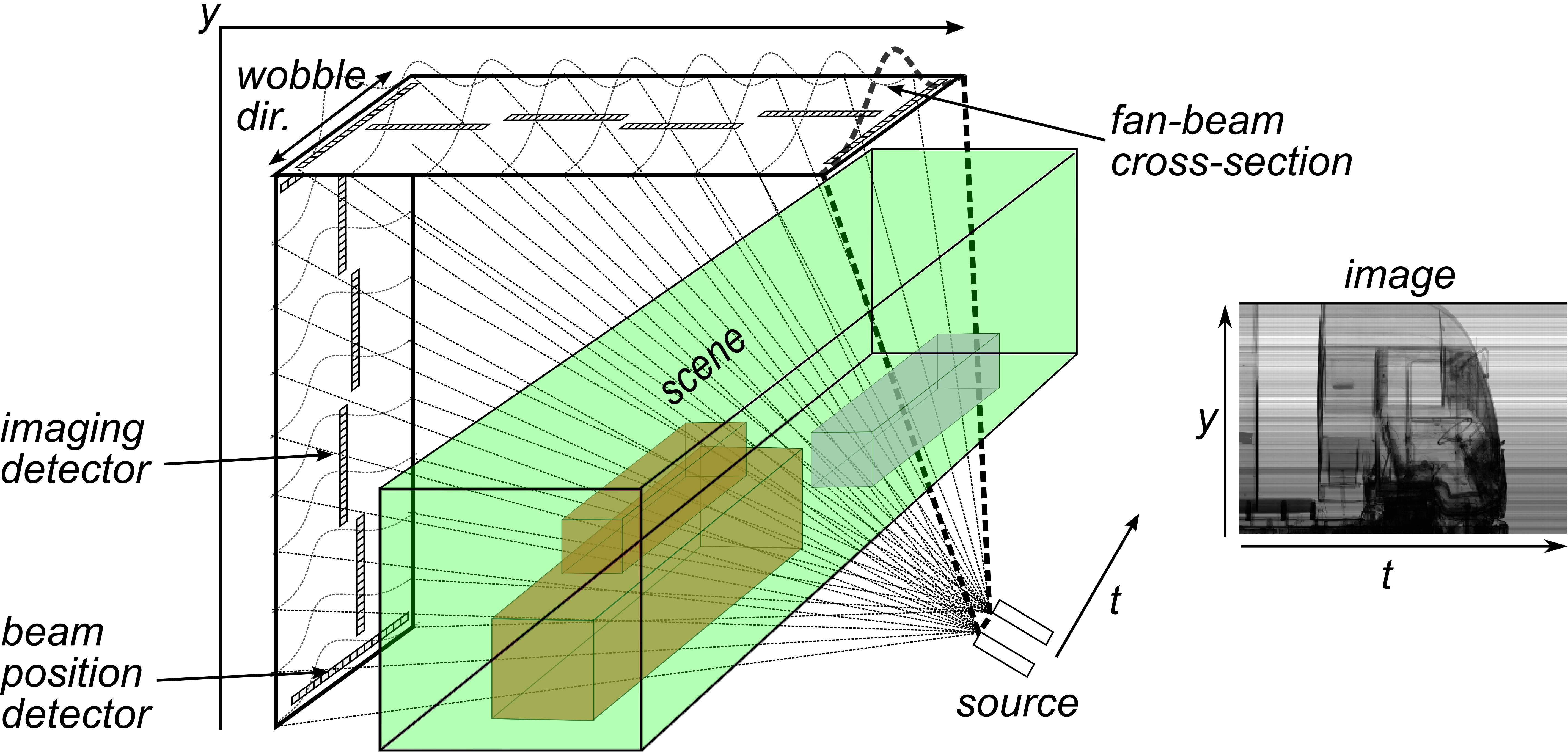}
     \caption{A typical transmission radiography system. Translation of the scene relative to the source and detector produces image columns, whilst each image row corresponds to a single sensor position in the imaging array. The set-up, considered in this paper, has been modified by addition of four {B}eam {P}osition {D}etectors {(BPDs)} which are are detector strips oriented at $90^\circ$ to the imaging array. These allow the intensity profile across the beam width to be measured.}
  \label{fig:G60}
\end{figure}

{In our previous work~\cite{Rogers2014} we proposed that wobble can be measured using Beam Position Detectors (BPDs) which are placed perpendicular to the imaging array (Fig.~\ref{fig:G60}). Wobble was estimated by performing Gaussian model fitting to the BPD data to obtain instantaneous beam position estimates. These instantaneous estimates were Bayesian fused with an estimate from an Auto-Regression (AR) to make estimates more robust for scanning moments where the BPD was non-uniformly obscured by an object in the scanned scene. The wobble estimates were then used to make corrections to air-only images in order to quantify performance. We determined that we could correct 70\% of image error due to wobble.}

{In this work, we follow a similar approach using BPDs but with several new contributions: (i) a model of image formation in the presence of wobble and other scanner design imperfections such as variable imaging sensor misalignments, variable sensor responses, and source fluctuation; (ii) improved wobble estimation using a Random Regression Forest (RRF) model for improved instantaneous estimation of wobble and its uncertainty, (iii) improved image correction by estimating the relative offsets of sensors, and (iv) estimation of sensor sensitivities by Sum of Squared Error (SSE) minimisation model fitting. Furthermore, we extend testing of image correction methods to include qualitative evaluation on images of complicated scenes.}

{In the next section we set out the technical background and }review related work. In {Sec.~\ref{sec:artefactOrigins}, we give a} precise description of the effects of wobble and {a method of image correction based on using BPDs to estimate fixed (e.g. sensor sensitivities, sensor offsets, and beam geometry) and dynamic (e.g. wobble and photon flux) system parameters}. In Sec.~\ref{sec:wobbleEstimation} we propose a method for estimating {the} dynamic system parameters. {Finally, i}n Sec.~\ref{sec:results}, we test these methods on images {that we have collected} from a Rapiscan Eagle\textregistered G60, a large-scale transmission X-ray gantry system, modified by the addition of four BPDs.

\section{{Background and }Related work}
\label{sec:relatedWork}

{Large-scale transmission radiography scanners operate either in portal or traverse mode, and are sometimes capable of both~\cite{reed2008throughput}. In portal mode the scanner is stationary and the scene moves between the source and imaging array at a controlled speed. In traverse mode the detector and source move either side of the stationary scene. Portal mode is most useful in high-throughput scenarios; vehicles can drive through the scanner arch without the driver having to exit the vehicle or a rail-scanner can scan multiple cargo containers carried by train at up to 60 km/h~\cite{Rogers2015}. Traverse mode is useful in security scenarios where an unoccupied vehicle cannot be interfered with, such as if it suspected to be a car- or truck-bomb, or if it needs to be covertly inspected so as not to raise suspicions. The traverse mode is also useful for scanning lines of stationary cargo containers at ports~\cite{reed2008throughput} The traverse mode has advantages in some cases: (i) the scanned vehicle is unoccupied, so higher doses can be used, resulting in higher precision images; (ii) there is greater control over scanning speed and detector-object distance resulting in less spatial warping of the captured image; and (iv) they have a compact scanning footprint~\cite{orphan2005advanced}.} 

\begin{figure}[htbp]
   \centering
\includegraphics[width=0.9\textwidth]{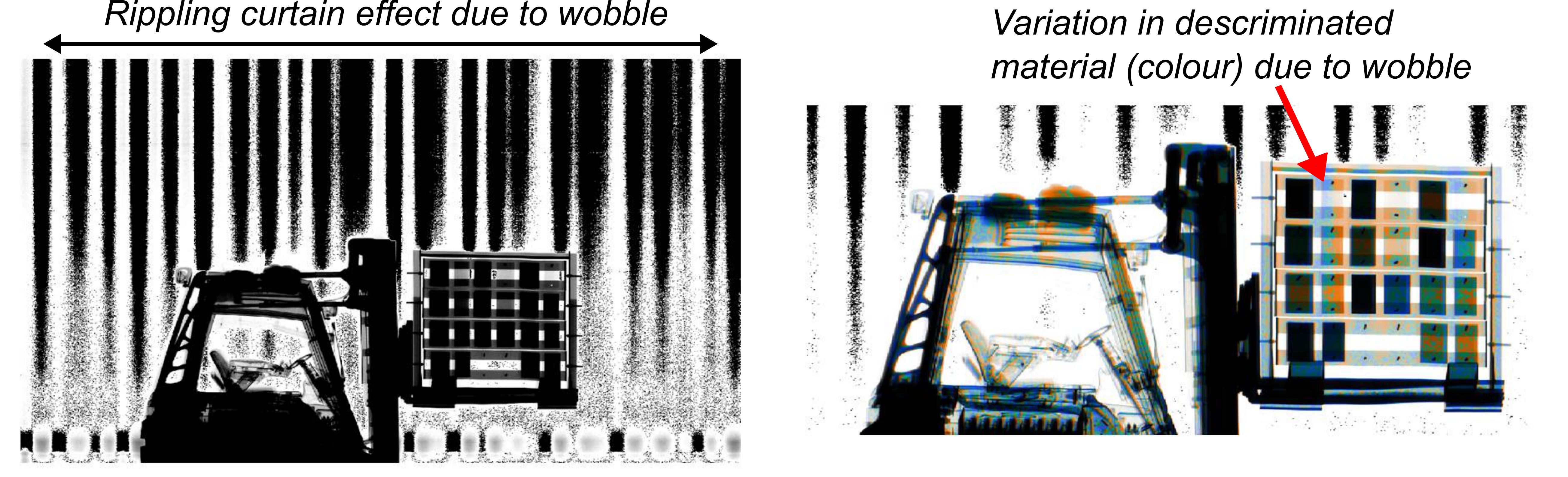}
     \caption{An X-ray image of a fork-lift truck from a mobile scanner with mechanical instability (left) and the same image with material discrimination applied (right). The image grey-levels have been windowed to make visible the small changes in image value due to wobble. {Wobble leads to a rippling curtain effect across the image}. Each rectangular test piece corresponds to a single material of uniform thickness. The wobble artefact effects the classification of material type; the classification of a single test piece can change from plastic through to steel due to wobble. {This results in a colour change across the test piece (indicated by red arrow) in the material discrimination image, where there should be no change}.}
  \label{fig:wobble}
\end{figure}

In the traverse mode, the imaging array may wobble as it moves across the scene due to uneven ground or vibrations from the engine (truck systems), oscillations in the boom (truck and rail systems), or due to wind or vibrations from traffic (truck and rail systems). This has a particular impact when operators search for threats placed in dense scenes, since under intensity windowing~\cite{Chen2005a} the wobble artefact becomes apparent (Fig.~\ref{fig:wobble}). Furthermore, wobble reduces the quality of material separation images~\cite{Ogorodnikov2002} since their computation is dependent on precise values. Discrimination of high atomic numbers is particularly important as it can reveal smuggled nuclear materials, or their shielding~\cite{orphan2005advanced,chen2007dual}. Wobble occurs in both truck-mounted and gantry systems. In truck-mounted systems wobble is variable from scan to scan, but in gantry systems it is systematic. In this work we study a gantry system, since it allows determination of the wobble ground truth, but our methods can equally be applied to truck-mounted systems.

{To our knowledge, other than our previous work~\cite{Rogers2014}, there have been no publications on addressing wobble in large-scale transmission radiography. However, wobble leads to artefacts in a range of imaging devices, including micro-CT and C-arm CT. We describe the most relevant work here and how it relates to the wobble effect that we attempt to measure and correct in this work.}

{C-arm CT systems suffer from wobble as the gantry rotates. This means that individual projections are translated relative to those captured by an wobble-free ideal device. Authors note that the wobble of the C-arm gantry is often repeatable over periods of up to two years and so wobble artefacts can be corrected by a one-off system calibration~\cite{silver2000determination,fahrig2000three}. This is similar to some large-scale transmission systems, particularly those that are in fixed deployment and the gantry moves along rails, where the wobble effect tends to be systematic. However, in truck-mounted mobile systems wobble is much more unpredictable due to variable scan speed and due to variations in the topology of the surface that the truck traverses. Moreover, in C-arm CT wobble artefacts tend to lead to a blurring effect in the reconstructed image due to the misalignment of individual projections, whereas in large-scale transmission systems, wobble mostly leads to image intensity variations as the fan-beam comes in and out of alignment with the detectors. Indeed, geometric image distortions can be observed if wobble is particularly severe, but this will be the focus of later work.}

{Silver \emph{et al.}~\cite{silver2000determination} propose a method for determining and correcting wobble in C-arm CT. The authors assume that the wobbling motion of the C-arm is the identical for each image capture process, and so calibrate wobble correction based on a phantom image. The phantom consists of a helical structure of tungsten carbide spheres (pellets). The calibration computes wobble coefficients that are used directly in image reconstruction to obtain a wobble-artefact free image. The wobble coefficients are determined by fitting a mapping from physical space to projection space using least-squares. Fahrig and Holdsworth~\cite{fahrig2000three} also adopt a calibration approach to determine projection translations. They use a bi-cubic spline interpolation to determine translated projections. Since the calibration process determines translations at discrete gantry angles, they linearly interpolate between them to obtain estimates for different projection angles if required.}

{Wobble is also observed in micro-CT systems, but the wobble manifests in the rotation table since the detector and source are kept stationary~\cite{sasov2008compensation}. In this case, wobble again leads to a blurring effect in the image, quite different to the effect observed in large-scale transmission systems. Authors have investigated image-based, calibration and online methods to correct for wobble.}

{Sasov \emph{et al.}~\cite{sasov2008compensation} investigate and evaluate an image-based and a calibration-based method. The image-based method is an iterative compensation scheme, which first does an initial reconstruction using filtered back-projection, yielding blurry wobble affected reconstruction. Estimates for projection translations to compensate for wobble are determined by comparing original projections with corresponding forward-projected image estimates. The comparison is done either by cross-correlation or least-squares. Under these translations a new reconstruction is made and the process is iterated until the reconstructed image is satisfactory. The calibration-based method, measures wobble in a short reference scan directly before or after image capture to determine the compensatory translations of individual projections. They measure the position of the focal spot, relative to a metal pin placed in the scene, by fixing a fine metal mesh to the X-ray source. The authors claim that the second method is more suitable for slow drifts (wobble) and that the approach is faster and less computationally demanding than the iterative based method. However, the image-based method has the advantage of working purely on measured image values.}

{Zhao \emph{et al.}~\cite{zhao2016method} Propose an online method, which uses capacitive distance sensors to measure the wobble of the rotation table in Micro-CT. The measurements are used to translate individual projections to compensate for the displacement of the rotation table due to wobble. The authors report that the methods improve images by 53.1\% and 65.5\% when calibrating projections in the horizontal and vertical directions, respectively.}

{Due, to the unpredictable component (e.g. wind, uneven topology, vibrations) of wobble in large-scale transmission radiography, it is not possible to correct wobble purely by calibration. Image-based methods (without using BPDs or prior knowledge about large-scale radiography), such as Total Variation (TV) denoising or Translation Invariant Wavelet Shrinkage (TIWS)~\cite{mouton2013evaluation}, may be applicable, however they are difficult to use in practice without prior information on the severity of the wobble artefact which we measure (online) in this work. In this contribution, we use both a calibration procedure and an online method. The calibration procedure is used to estimate a number of parameters that are fixed for a given system, including: misalignments of imaging sensors; the collimated width of the fan-beam; and the sensitivities of individual sensors due to housing attenuation and their intrinsic response. The online component, is for the estimation of wobble and estimation of the fluctuation in the photon flux, which can both vary unpredictably during a traverse mode scan. We describe these methods in the next section.}

\section{A model of image formation with wobble}
\label{sec:artefactOrigins}
\begin{figure}[htbp]
   \centering
\includegraphics[width=0.9\textwidth]{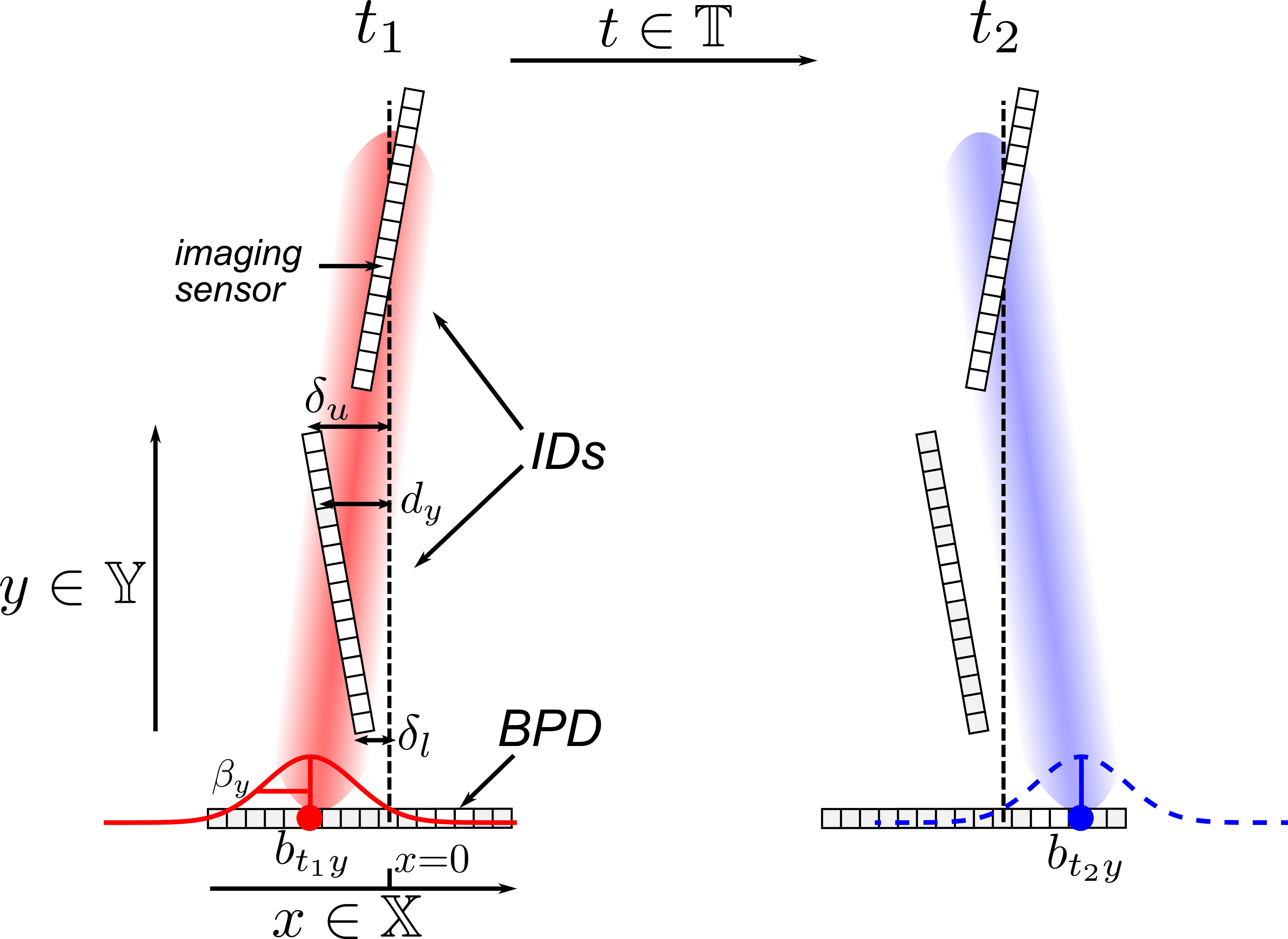}
     \caption{\emph{Left}: Part of the imaging array showing two misaligned {I}maging {D}etectors (IDs), a {B}eam {P}osition {D}etector (BPD), and a wobbling fan-beam. The magnitude of the wobble and the sensor misalignments have been exaggerated in this figure. The offsets $d_y$ for individual imaging sensors are confined to a linear function determined by the offsets $\{\delta_l,\delta_u\}$ of the ID endpoints. The fuzzy bars illustrate the fan-beam incident on the imaging array. The Gaussian (width $\beta_y$ and position $b_{ty}$) shows the profile of the fan-beam on the BPD. \emph{Right}: A later time point $t_2 > t_1$. Due to wobble, the fan-beam has moved relative to the imaging array so that the intensity recorded by the imaging detectors has changed. This leads to an effective loss of image precision. Correction requires estimation of the beam displacements $b_{ty}$ and the offsets $d_y$ to be estimated. The $b_{ty}$, dense in $t$ and $y$, can be interpolated from estimates dense in $t$ but computed at the sparse $y$ values where BPDs are located.}
  \label{fig:BPDSchematic}
\end{figure}

{To describe image formation with a wobbling detector, we use three coordinate systems (Fig.~\ref{fig:BPDSchematic}). We denote: the coordinates of imaging sensor pixels along the $\Gamma$-shaped imaging array (image vertical) by $y \in \mathbb{Y}$; the time coordinates indexing each scanning moment during image acquisition (image horizontal) by $t \in \mathbb{T}$;  and the coordinates along the orientation of the BPDs (perpendicular to the beam and imaging array) by  $x \in \mathbb{X}$. The origin $x=0$ is taken as the vertical midline (dashed in Fig.~\ref{fig:BPDSchematic}) of the imaging array.}

{The formation process of an image $I_{ty}\in \mathbb{R}^+$ is described as follows. The X-ray source emits a photon flux $A_{t}\in \mathbb{N}$ at scanning moment $t$. This flux is collimated into a fan-beam of width $\beta_y$, which has a spatial distribution on the imaging plane according to}
\begin{equation}
\exp{\left(-{(b_{ty} - d_y)^2}/(2\beta_y^2)\right)}.
\end{equation}
{The parameters $b_{ty} \in \mathbb{X}$ define the displacements of the beam cross section maximum from the vertical midline: when wobble occurs this varies with $t$ and $y$, without wobble only with $y$. The parameters $d_y\in \mathbb{X}$ are the horizontal offsets of the imaging sensors from the vertical midline. For a given linear ID with endpoint offsets $\{\delta_l,\delta_u\}$ we constrain $d_y$ to a linear function}
\begin{equation}
\label{eq:linearConstraint}
d_y := (y_u-y_l)^{-1}((y_u-y)\delta_l + (y-y_l)\delta_u) ,\ \text{where}\ y_l<y<y_u.
\end{equation}

{The X-ray photons pass through the scene and interact via absorption and scattering, and we denote the scene transmission by $S_{ty}\in [0,1]$. This is dependent on the thickness and type of material composing the scene. The final measured image is determined according to a sensitivity factor $R_y\in [0,1]$, which incorporates (i) the fraction of photons that are transmitted through the sensor housing and not absorbed or scattered, and (ii) the fraction of photons impinging on the detector that are counted (the intrinsic response of a sensor).}

{Therefore, the final image,} assuming no cross-pixel effects such as photon scatter or detector cross-talk, is approximated by

\begin{equation}
\label{eq:imageEqn}
I_{ty} = A_t \cdot \exp{\left(-{(b_{ty} - d_y)^2}/{(2\beta_y^2)}\right)} \cdot S_{ty} \cdot R_{y}.
\end{equation}

The scene transmission $S_{ty}$ is the physical quantity that we are trying to measure, therefore the ideal image is
\begin{equation}
\label{eq:idealImageEqn}
\underbrace{S_{ty}}_{\text{ideal}} = \underbrace{I_{ty}}_{\text{raw}}\cdot \underbrace{(A_t \cdot \exp{\left(-{(b_{ty} - d_y)^2}/{(2\beta_y^2)}\right)}  \cdot R_{y})^{-1}}_{\text{correction factor}}.
\end{equation}

To obtain the ideal image, one must estimate the different components of the correction factor. In the \emph{portal} scanning mode, correction is straightforward. Absence of wobble means that $b_{ty} = b_y$, so that all that needs to be dealt with is:
\begin{enumerate}
\item image column variations due to fluctuations of the photon source $A_t$;
\item image row variations due to sensitivity $R_y$, and the fixed position and geometry of the beam $\exp{\left(-{(b_{y} - d_y)^2}/{(2\beta_y^2)}\right)}$;
\item image pixel variations due to Poisson variation in the number of photons that reach an imaging sensor.
\end{enumerate}
The image column and row variations (1 \& 2) can be corrected by normalising the columns and rows in the image respectively. In this work we do not attempt to correct for Poisson variation (3), however there are several denoising algorithms for Poisson-distributed noise~\cite{Cheng2015,Leftkimmiatis2009} in the literature. Note that Poisson variation can also be ameliorated by increasing the beam intensity or exposure time, but this has implications on safety and cost.

In the \emph{traverse} scanning mode, where wobble \emph{does} occur, the correction is complicated. The beam position $b_{ty}$ now varies with $t$ as well as $y$, and the imaging sensor offsets $d_y$ must now also be estimated. These and the other parameters in the correction factor (Eq.~\ref{eq:idealImageEqn}) can be separated into two classes; (i) \emph{system parameters} ($\beta_y$,$\{\delta_l,\delta_u\}$ and $R_y$) that are estimated in a one-off calibration which we describe {below}, and (ii) \emph{dynamic parameters} ($b_{ty}$, $A_t$) that are estimated per time point (online). The source variation $A_t$ is straightforward to address by taking an image patch from a single ID close to the source and averaging it over rows. In the remainder of this paper we work on $A_t$-corrected images. In Sec.~\ref{sec:wobbleEstimation} we describe a method to estimate $b_{ty}$.

In the one-off calibration, for each BPD we estimate $\beta_y$ (beam width at the BPD location) and $R_x$ (the sensitivity of the sensors along the BPD). For each ID, we estimate $\{\delta_l,\delta_u\}$ (the misalignments of the ID at its endpoints) and $R_y$ (the sensitivity of the sensors along the ID). {The estimates} are determined by model fitting to data collected during a traverse (wobbling) scan of an air-only scene. Although wobble has a detrimental effect on image precision, we benefit from wobble in these estimations since it allows us to disentangle (i) $\beta_y$ and $R_x$, and (ii) $b_{ty}$ and $\{\delta_l,\delta_u\}$.

{The calibration is two-step and summarised as follows. First, we perform a Sum of Squared Errors (SSE) minimisation model fit using a Gaussian model of the fan-beam incident on the BPD, masked by the sensor sensitivities. In the fitted model, the Gaussian centre is allowed to vary freely with time but the beam width and sensitivities are unvarying. Having estimated the unvarying beam widths and the time-varying beam positions at each BPD, we linearly interpolate these to the positions of the sensors of the IDs. With these estimated, next we model fit to determine the ID parameters. We perform a SSE fit to the data from each ID to jointly estimate $\{\delta_l, \delta_u\}$ and $R_y$. The SSE is taken between the ideal image (raw image multiplied by correction factor, as in Eq.~\ref{eq:idealImageEqn}) and a uniform unit-valued image. The correction factor is composed using the interpolated $\beta_y$ and $b_{ty}$ estimates (from step 1), and the estimated parameters $\{\delta_l,\delta_u\}$ and $R_y$.}

\section{Wobble estimation algorithm}
\label{sec:wobbleEstimation}
To estimate wobble for inhomogeneous scenes, we need to estimate $b_{ty}$ at the BPDs, and then interpolate it along the IDs. However, the simple model fitting of the previous section is not applicable for inhomogeneous scenes. At some scanning moments the beam will be distorted from a Gaussian shape, and at other moments it will be undetectable due to dense loads. To cope with this, we estimate the beam position at time $t$ by fusing an instantaneous estimate $\hat{b}_{\text{inst}}$ (with uncertainty $\hat{\sigma}_{\text{inst}}$), with an estimate $\hat{b}_{\text{prior}}$ (with uncertainty $\hat{\sigma}_{\text{prior}}$), based on the previous $n$ beam position estimates. 

\subsection{Instantaneous estimation}
\label{sec:instantaneousEstimation}
\begin{figure}[tb]
   \centering
\includegraphics[width=0.9\textwidth]{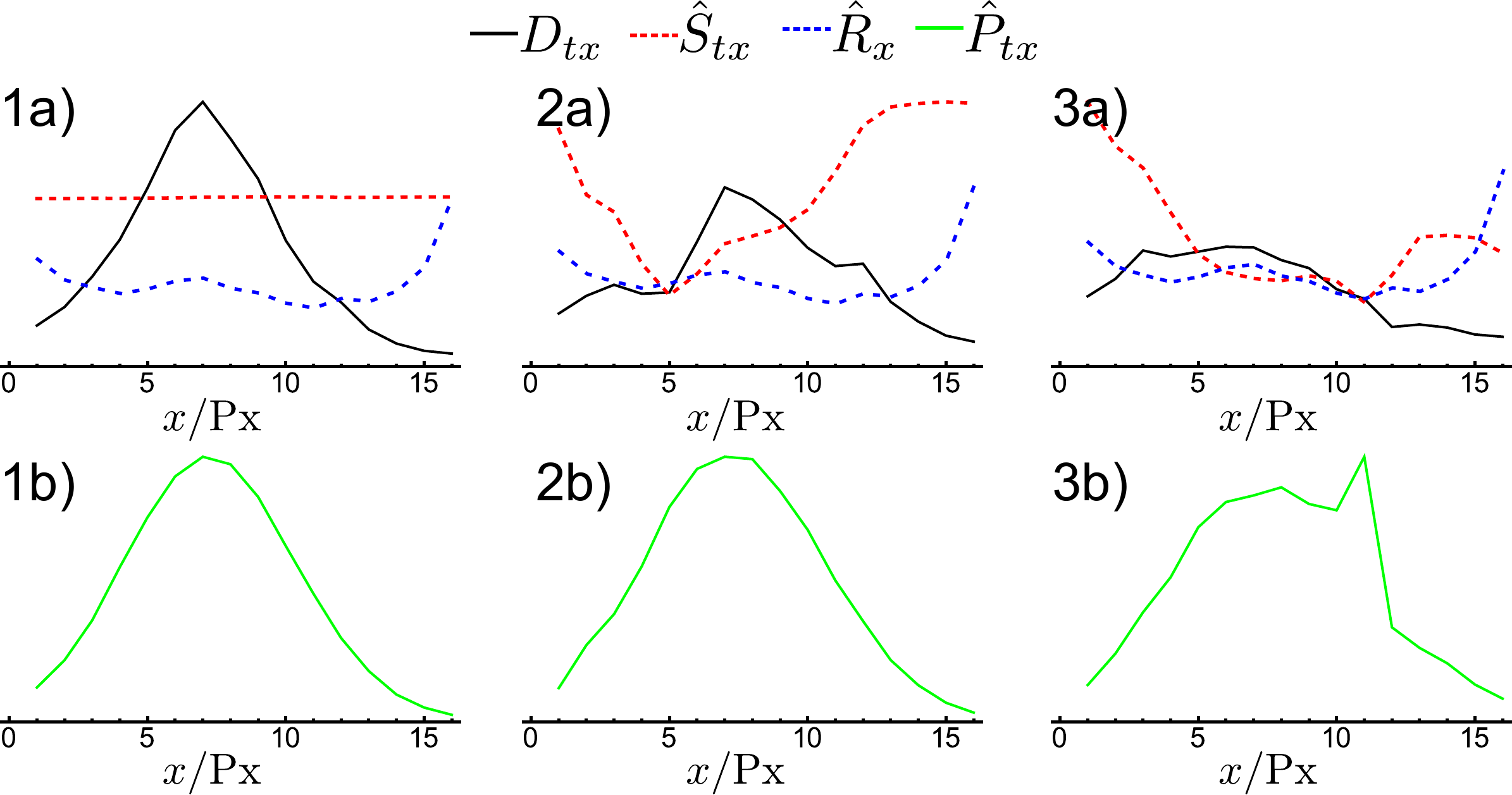}
     \caption{Examples of the estimated beam profile $\hat{P}_{tx}$ (green) computed by dividing the measured {Beam Position Detector (}BPD{)} profile $D_{tx}$ (black) by estimates of the scene transmission $\hat{S}_{tx}$ (red) and the sensor sensitivity $\hat{R}_{x}$ (blue). \emph{Left:} Example of a homogeneous scene, and thus the estimate $\hat{S}_{tx}$ is flat, resulting in a Gaussian $\hat{P}_{tx}$. \emph{Middle:} Example of an inhomogeneous scene, the resulting $\hat{P}_{tx}$ is approximately Gaussian. \emph{Right:} Example of a dense inhomogeneous scene, where the resulting $\hat{P}_{tx}$ is non-Gaussian which we attribute to photon scatter.}
  \label{fig:beamEstimates}
\end{figure}

The profile ($D_{tx}$) measured at each instant by a BPD, is a multiplicative combination of (i) the beam profile ($P_{tx}$), (ii) the scene transmission ($S_{tx}$), and (iii) the sensitivity ($R_x$). We estimate the beam profile from the measured profile, fixed estimates of the sensitivity (Sec.~\ref{sec:artefactOrigins}), and dynamic estimates of the scene transmission estimated from previous time-points of the BPD signal, according to \begin{equation}
\label{eq:beamEstimates}
\hat{P}_{tx} = D_{tx}/(\hat{R}_x\, \hat{S}_{tx}).
\end{equation}
This estimation works well in cases where the scene is not too dense (Fig.~\ref{fig:beamEstimates}, 1b \& 2b); but when it is the estimated beam profile can be inaccurate due to (i) the low (noise-dominated) sensor signal, or (ii) deviation of photon trajectories due to scatter (Fig.~\ref{fig:beamEstimates}, 3b).

\begin{figure}[htbp]
   \centering
\includegraphics[width=0.9\textwidth]{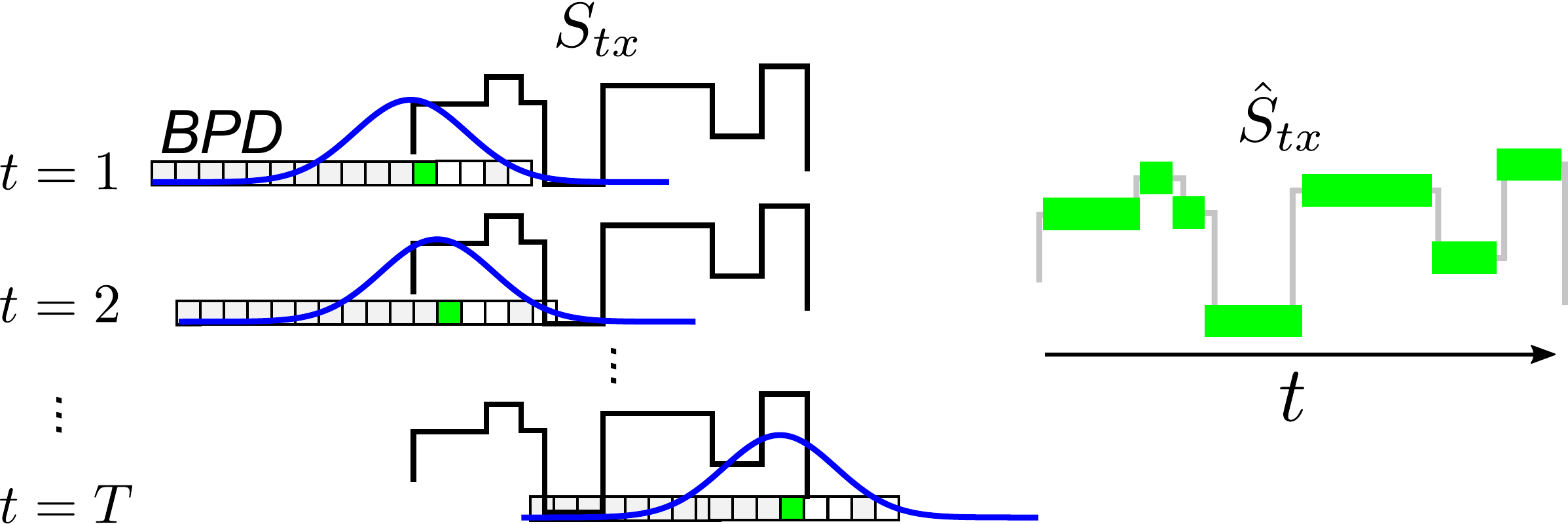}
     \caption{\emph{Left:} Illustration of a {Beam Position Detector (}BPD{)} being translated across a scene during a scan. At consecutive timepoints $t = \{1,2,\dots,T\}$ a given sensor (green) samples consecutive points in the scene. \emph{Right:} plotting these samples as a function of $t$ yields an estimate of the scene transmission. Each BPD sensor gives a similar estimate, and we perform a weighted average of them to reduce the noise in the final estimate $\hat{S}_{tx}$. Sensors towards the ends of the BPD, which receive low signal, are given a lower weighting in the average than those near the Gaussian centre which receive a higher signal.}
  \label{fig:sceneFunction}
\end{figure}

We estimate the scene transmission function $\hat{S}_{tx}$ using measurements of the BPD as it slides across the scene. A given pixel on the BPD samples each point, at its $y$-value, in the scene (Fig.~\ref{fig:sceneFunction}). Plotting the response of this pixel as a function of time gives an estimate of the scene transmission function. Since each of the BPD sensors also sample each point in the scene, we can construct a similar estimate for each sensor. The final estimate of $\hat{S}_{tx}$ is obtained by taking a weighted average of the estimates from each of the sensors. We take the weighted average to reduce noise in the estimate from sensors that are aligned with the low signal tails of the Gaussian cross section. 

{With the estimate of the beam profile ($\hat{P}_{tx}$), we can estimate the instantaneous beam position $b_{\text{inst}}$ and its uncertainty $\sigma_{\text{inst}}$. The estimator should be able to deal with non-linear relationships in the data and be able to produce data dependent uncertainty estimates. We have experimented with using Gaussian model fitting, as used in Ref.~\cite{Rogers2014}, but find that the non-normal distribution of the errors makes estimation of the uncertainty unreliable.}

{In this contribution, we use a Random Regression Forest (RRF)~\cite{breiman2001random} to construct a robust estimator of the beam position from the beam profile estimates. A RRF model is based on an ensemble of decision trees and is capable of modelling non-linear relationships as required. Each tree in the RRF produces an estimate of the beam position. We obtain estimates of the instantaneous beam position $\hat{b}_{\text{inst}}$ and its uncertainty $\hat{\sigma}_{\text{inst}}$ by taking the mean and standard deviation of the tree responses, respectively. We observe, for this study, that the standard deviation of the tree responses has a strong correlation with the actual error in the beam position estimate. Other advantages of RRF is that it is fast to train and deploy, and resistent to overfiting.}

{In the RRF, $N_t$ trees are constructed top-down with bagging and random subspace sampling. Internal nodes are split using standard thresholding, and optimised according to the Residual Sum of Squares (RSS). At each split $m$ features (i.e. BPD pixels; elements of $\hat{P}_{tx}$ at fixed $t$) are randomly sampled. For stopping criteria, we do not set a maximum tree depth and enforce a minimum of two samples per split. To tune the $N_t$ and $m$, we first set $m$ to the recommended default ($m=1/3 \times\text{\# features} = 5$) for regression. We then vary $N_t$ and assess the RMSE to choose a sufficient number of trees so that the RMSE is stabilised but not too many that computation time is slow. With the $N_t$ fixed, we then vary $m$ from 3 to 12 to find the optimal RMSE, before verifying $N_t$ again as before. By this method we determined that $N_t=500$ was adequate and the default $m=5$ was optimal.}

In this work we use the \texttt{randomforest-matlab} implementation of RRFs~\cite{Jaiantilal}. For training, the ground truth values of the beam displacement were obtained by use of a gantry system in traverse mode, described later in Sec.~\ref{sec:results}. We train a separate RRF for each BPD, using $1.4\times 10^5$ measurements from five independent scans so that there is no overlap with the test images used in Sec.~\ref{sec:results}.

\subsection{Estimation based on previous estimates}
\label{sec:AR}
\begin{figure}[htbp]
   \centering
\includegraphics[width=0.9\textwidth]{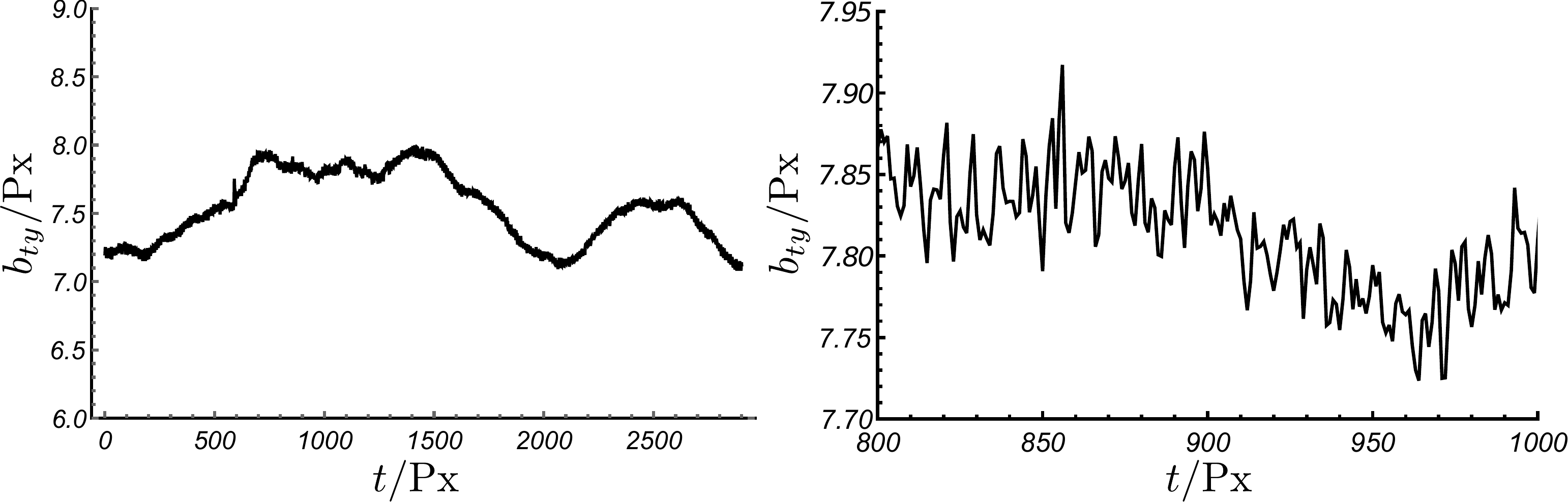}
     \caption{\emph{Left:} Beam position $b_{ty}$ as function of time $t$ during a wobbling air scan. \emph{Right:} A zoom so that the high frequency component of the wobble is visible.}
  \label{fig:bAirTrace}
\end{figure}

{In cases where the BPD is heavily obscured (low signal-to-noise), the RRF-based instantaneous estimate will give a poor estimate of the beam position and a high uncertainty. In these cases, we want the beam position estimate to be sensible, and to achieve this we incorporate information about prior beam positions using an Auto-Regression (AR). The wobble of the detector array is partly deterministic (consider a swinging pendulum), but also stochastic due to the variable scanning surface, wind and vibrations. An AR is capable of learning some of the deterministic wobble  whilst allowing for stochastic variation. It is also simple to implement and fast to compute. Additionally, we observe (Fig.~\ref{fig:bAirTrace}) that the beam position trace has a high frequency component due to fluctuations of the photon source, possibly originating from electronic circuitry; and a low frequency component due to the wobble of the imaging array. The high frequency element makes simple estimation, based on the previous time point, unreliable. An AR, however, allows incorporation of $n$ previous timepoints, where $n$ can be tuned on data to achieve best performance. Moreover, the AR approach effectively smooths out erroneous estimates from previous time-points, but is beneficial over other smoothing filters (e.g. median filter) since it is possible to propagate previous errors to be used in fusion (Sec.~\ref{subsec:fusion}) with the instantaneous estimate.}

The AR model predicts the current beam position based on a linear combination of the previous $n$ beam positions with an added, normally distributed, perturbation
\begin{equation}
b_{t} = \sum_{t'}w_{t'} b_{t-t'} + N(0,\epsilon^2)\; \text{s.t.}\; \sum_{t'} w_{t'} =1,
\label{eq:AR}
\end{equation} 
where $1 \leq t' \leq n$. 

The {Auto-Regression (AR) weights} $w_{t'}$ are determined by model fitting Eq.~\ref{eq:AR} to an independent air scan. The constraint $\sum w_{t'} = 1$ ensures that the model does not have an unrealistic systematic drift. The uncertainty $\epsilon$ is determined by applying the model to a second air-only scan and computing the {R}oot-{M}ean-{S}quare {E}rror (RMSE). The fitted model is used to generate the prior beam position estimate and its uncertainty according to:
\begin{equation}
\hat{b}_{\text{prior}} = \sum_{t'} w_{t'} \hat{b}_{t-t'},\ \hat{\sigma}_{\text{prior}}^2 = \sum_{t'} w_{t'} \hat{\sigma}_{t+t'}^2+\epsilon^2.
\end{equation}
Note that the uncertainties from previous timepoints are propagated when forming this estimate{, so that if the AR operates on previous estimates that are highly uncertain they are incorporated into the AR uncertainity, which is useful in the fusion step.}

\subsection{Fusion of estimates}
\label{subsec:fusion}
\begin{figure}[htbp]
   \centering
\includegraphics[width=0.9\textwidth]{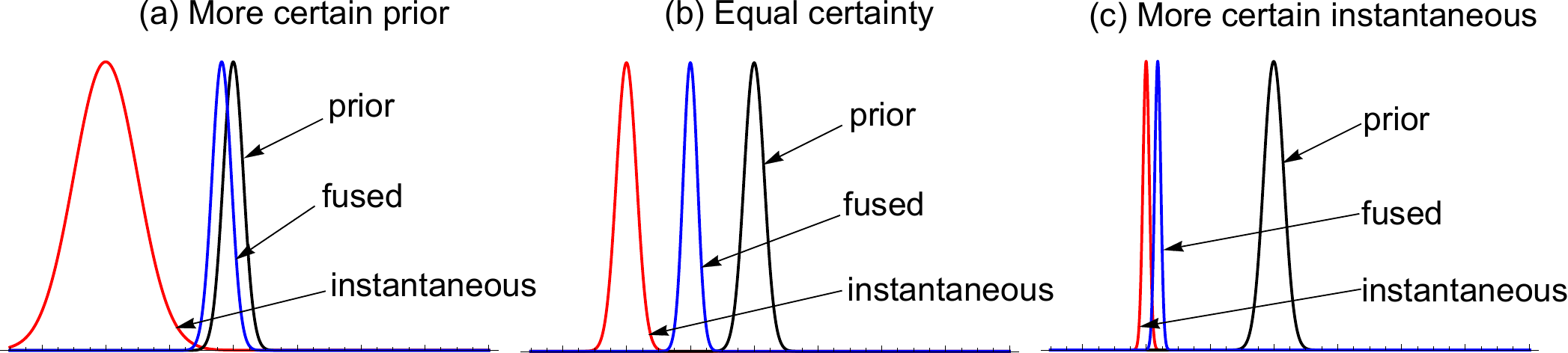}
     \caption{{Demonstration of Bayesian fusion of a prior estimate (black) and an instantaneous estimate (red) to obtain a fused estimate (blue). The width of the Gaussians correspond to the uncertainty on the estimate, and their centroid to the estimate value. The $x$-axis can be imagined as the Beam Position Detector (BPD). In (a) the prior has a higher certainty than the instantaneous estimate and so the fused estimate is weighted towards the prior, in (c) the opposite is true. In (b) both estimates have equal certainty and so the fused estimate compromises between the two.}}
  \label{fig:fusionIllustration}
\end{figure}

{To incorporate the information from the previous time-points, we fuse the estimates from the AR and RRF models according to their uncertainties. The fusion should weight the final estimate more towards the AR if the RRF-based estimate is more uncertain (e.g. due to low signal-to-noise). Equally, if the AR uncertainty is high, because many of the previous $n$ RRF-based estimates were also uncertain, but the next instantaneous estimate is very certain, then the fusion should weight more towards the RRF-based instantaneous estimate. To achieve, this we use a Bayesian fusion, which is equivelant to a Kalman Filter~\cite{Faragher2012}. This approach is illustrated in Fig.~\ref{fig:fusionIllustration}.}

To estimate the beam position $\hat{b}_t$ and its uncertainty $\hat{\sigma}_t$, we Bayesian fuse the instantaneous estimate $\hat{b}_{\text{inst}}$ and it uncertainty $\hat{\sigma}_{\text{inst}}$ (Sec.~\ref{sec:instantaneousEstimation}) with a prior estimate $\hat{b}_{\text{prior}}$ and its uncertainty (Sec.~\ref{sec:AR}). This is expressed as:
\begin{equation}
\hat{b}_t = (\hat{b}_{\text{inst}}\hat{\sigma}_{\text{prior}}^2 + \hat{b}_{\text{prior}}\hat{\sigma}_{\text{inst}}^2)/(\hat{\sigma}_{\text{prior}}^2+\hat{\sigma}_{\text{inst}}^2),\ \text{with} \ \hat{\sigma}_t^2 = (\hat{\sigma}_{\text{prior}}^2\hat{\sigma}_{\text{inst}}^2)/(\hat{\sigma}_{\text{prior}}^2+\hat{\sigma}_{\text{inst}}^2).
\end{equation}

This weights the two beam position estimates by their uncertainty. If the uncertainty of an estimate is low then that estimate contributes more to the fused estimate. In particular, if the instantaneous estimate is uncertain because of dense shielding, the prior estimate will be relied on; but when it is certain it will dominate the overall estimate.

\section{Results}
\label{sec:results}
{For the purposes of this study, and to test out our methods, we} collected data using a modified Rapiscan Eagle\textregistered G60 transmission X-ray scanner. We rotated four of the IDs by $90^\circ$ to become BPDs. {The BPDs were placed at the extremes of the vertical boom and the horizontal boom, so that there were two BPDs per boom. The wobble characteristics are different at each location, for example wobble is most severe at the bottom of the vertical boom. Note that in a commercial implementation of BPDs, the system would have a full set of IDs with additional detectors for BPDs, but we have adopted this modification in experiments to reduce cost.} We collected air-only images in portal and traverse modes, and several traverse mode scans of objects (e.g. trucks, forklifts, scissor lifts) were performed. The scanner operates at $90\,\text{Hz}$ and has a pixel size of $5.6\,\text{mm}$, giving an effective spatial resolution of roughly $3\,\text{mm}$. The scanner uses a Bremsstrahlung beam with a cut-off energy of $6\,\text{MeV}$. {This is the same energy used in commercial systems, and gives enough penetration to achieve reasonable signal-to-noise ratio on the BPD for most objects.}

We adopted a gantry set-up, since it provides a ground truth for wobble. Wobble is observed in both gantry and truck-mounted systems, with a similar amplitude and frequency composition. However, for a gantry system, wobble is the same (modulo alignment) for each scan, but variable for truck-mounted systems. The gantry system allows us to obtain an accurate ground truth by aligning wobble estimates from an air-scan with the air parts of an object scan.

\subsection{System parameter estimation}
\begin{figure}[htbp]
   \centering
\includegraphics[width=0.9\textwidth]{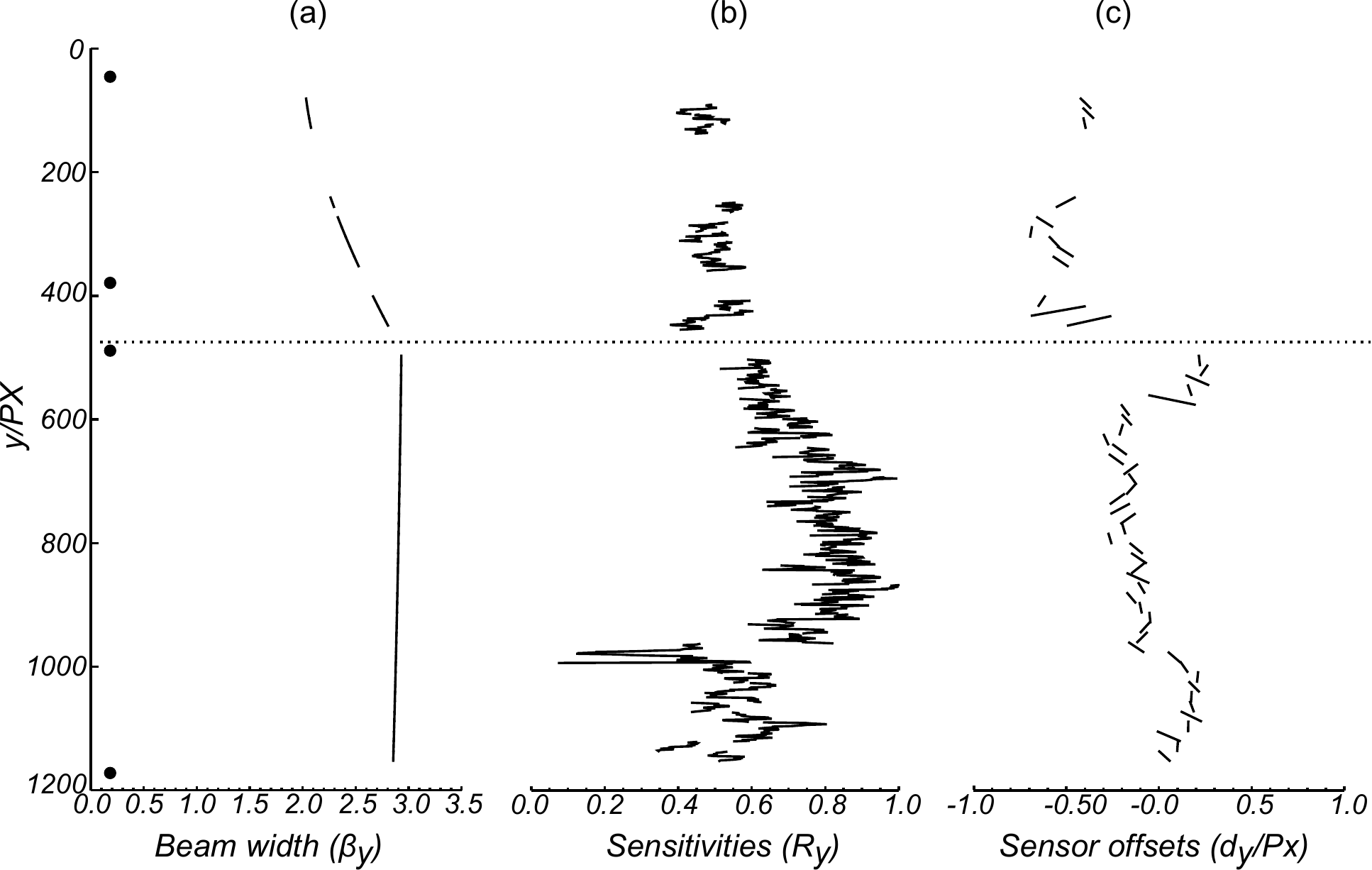}
     \caption{Estimated system parameters: {(a) beam width, $\beta_y$; (b) sensor sensitivities, $R_y$; and (c) horizontal imaging sensor offsets from the vertical, $d_y$}. The dashed horizontal line marks the transition from the vertical (below) part of the $\Gamma$-shaped imaging array, to the horizontal (above). The black dots indicate the $y$-positions of the {Beam Position Detectors (}BPDs{)}. Gaps in $y$-values are where an {Imaging Detector (}ID{)} has been removed or rotated to form a BPD in the experimental set-up. {The beam width increases (decreases) as the distance from source to the array increases (decreases), due to dispersion. The sensitivities fluctuate between adjacent sensors due to their different intrinsic responses. The estimated sensor offsets are piece-wise linear because they are grouped by ID, and are of the order of a few mm which is within the manufacturing tolerance of a system of this scale.}}
  \label{fig:estimateParameters}
\end{figure}

The system parameters $\beta_y$ {(beam width)}, $R_y$ {(sensitivities)} and $d_y$ {(imaging sensor offsets)} were estimated according to Sec.~\ref{sec:artefactOrigins}, and are shown in Fig.~\ref{fig:estimateParameters}. Small and large $y$-values correspond to the bottom and top of the image, or the vertical and horizontal parts of the $\Gamma$-shaped imaging array, respectively. The gaps in $y$-values are where an ID has been removed or rotated to form a BPD.

The estimate of $\beta_y$ {(Fig.~\ref{fig:estimateParameters}.a)} increases as you go along the horizontal of the $\Gamma$-shaped imaging array and away from the source due to beam dispersion; it then decreases as you go along the vertical of the $\Gamma$-shaped imaging array and slightly closer to the source. The sensitivities $R_y$ {(Fig.~\ref{fig:estimateParameters}.b)} have a lot of variation between adjacent imaging sensors due to their intrinsic response and due to variations in the housing of the $\Gamma$-shaped array. The estimated offsets of the IDs {(Fig.~\ref{fig:estimateParameters}.c)} are of the order of a few mm, which when compared to their $10\,\text{cm}$ length is plausible for a human engineer placing them during the construction of the scanner{, and is indeed within the manufacturing tolerance of a scanning device of this scale ($6\,\text{m}$ tall)}. {Note that the piecewise-linear nature of $d_y$ is due to the linear constraint places on each ID (Eq~\ref{eq:linearConstraint}).}

\subsection{Wobble estimation}

\begin{figure}[htbp]
   \centering
\includegraphics[width=0.9\textwidth]{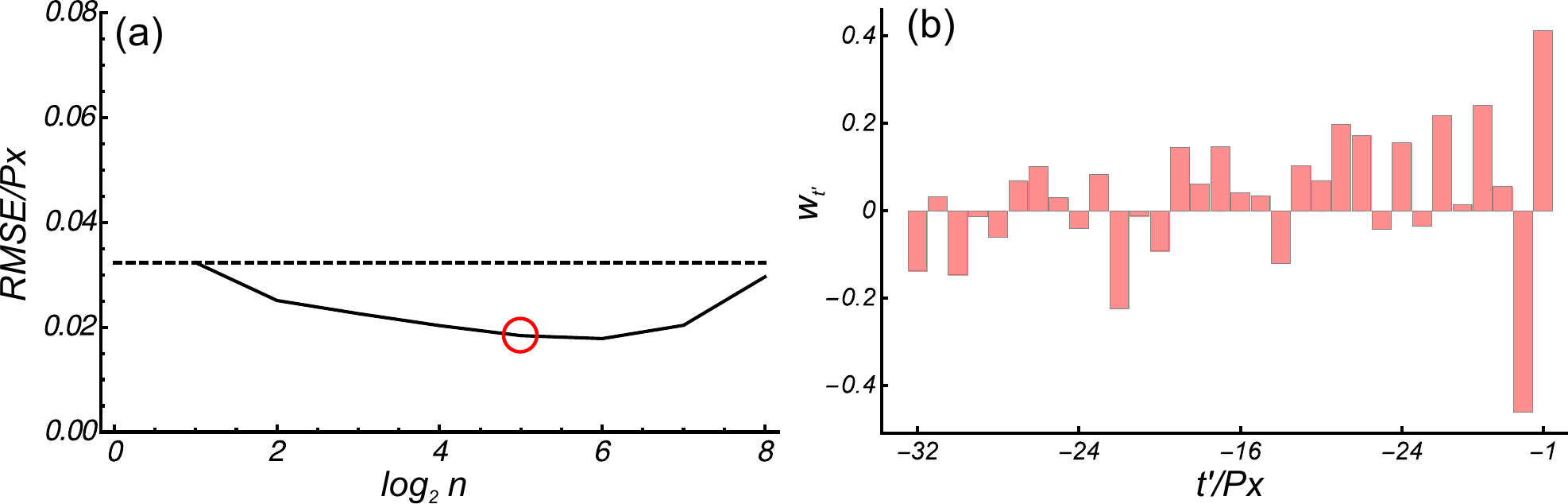}
     \caption{Auto-Regression (AR) model fit: (a) The {Root-Mean-Square Error (}RMSE{)} performance of the AR for different numbers of previous time-points $n$ included in the model (dashed line indicates the standard deviation of the beam position, red circle indicates the near-optimal $n=32$); and (b) the learnt AR weights $w_{t'}$ when $n=32$. {The RMSE decreases as the number of time points $n$ included in the model increases, it reaches an optimum at around $n=64$, before rising again due to overfitting. The AR weights have larger magnitude for the most recent time-points ($t'=-1,-2$) as expected, because these are most informative for predicting the next beam position. The oscillating structure in the weights is the AR's way of coping with the high frequency wobble component.}}
  \label{fig:AR}
\end{figure}

The AR was trained on an air-only traverse mode image. Fig.~\ref{fig:AR}(a) shows the RMSE performance of the trained AR on an independent air-only test image as a function of the number ($n$) of previous timepoints considered. As $n$ is increased the RMSE decreases, reaching a minimum at $n=64$, before the RMSE begins to grow. When $n$ gets too large the model overfits and performance deteriorates. We choose $n=32$ since the RMSE is near optimal but requires half the number of parameters. The AR weights for $n=32$ are shown in Fig.~\ref{fig:AR}(b). It shows that more importance is placed on the most recent $b$ estimates as expected. The oscillating structure is the AR system's way of coping with the high frequency component of the beam movement.

To assess the performance of the proposed beam position estimates, we test performance on ``easy'', ``intermediate'' and ``difficult'' scenarios from the collected data. For each, we compare the new RRF-based method for instantaneous estimation to the old Gaussian-based method from Ref.~\cite{Rogers2014}. We also compare the RRF-based method instantaneous method, with the fused estimate which we refer to as RRF-AR.

\begin{figure}[htbp]
   \centering
\includegraphics[width=0.9\textwidth]{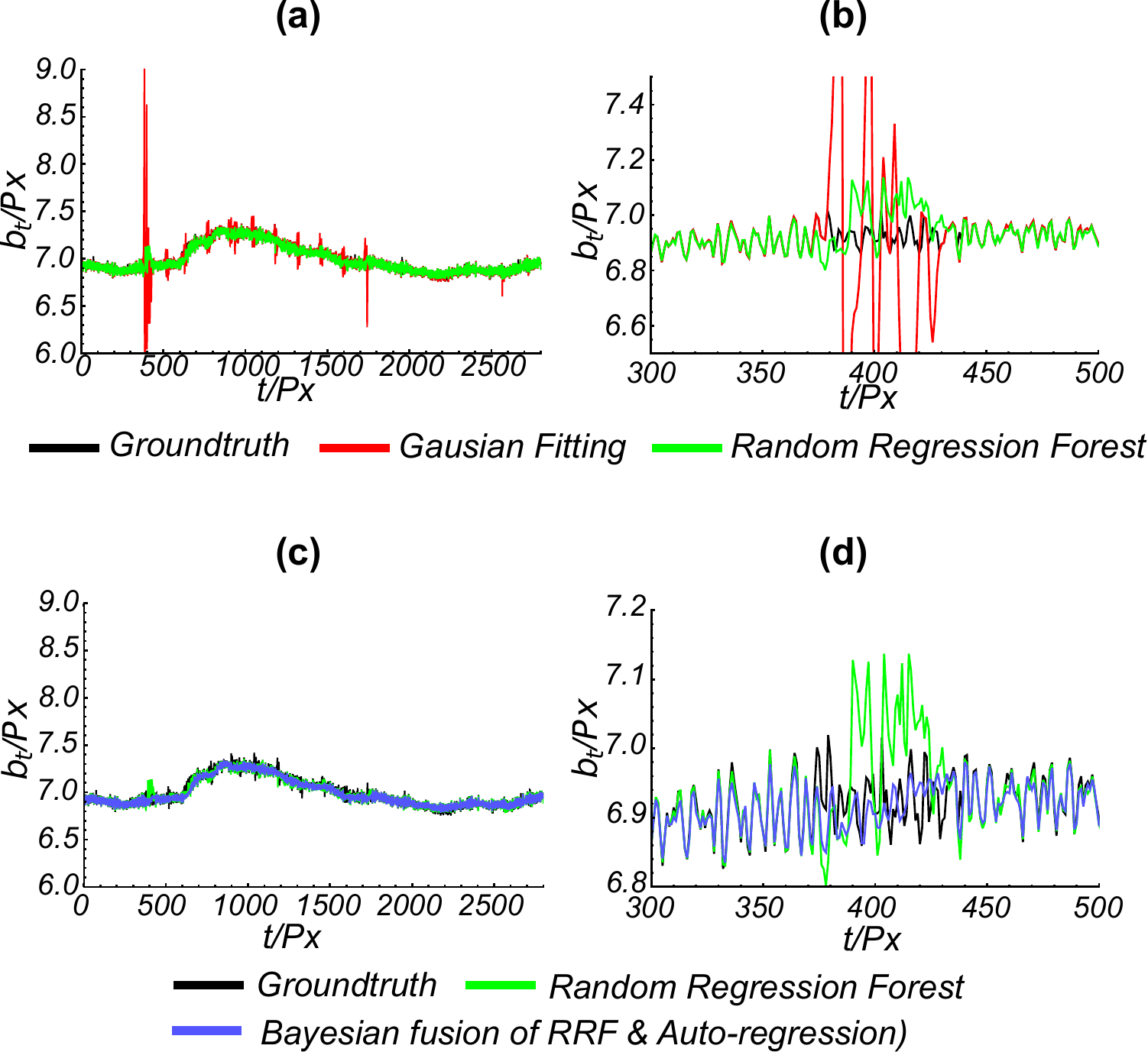}
     \caption{{Beam position estimates for the ``easy'' scenario. In (a) the new Random Regression Forest (RRF) based method (green) for instantaneous estimation is compared to the old Gaussian-based method (red), and the groundtruth (black). In (c) the RRF-based method (green) for instantaneous estimation is compared to its Bayesian fusion with an Auto-Regression (RRF-AR; blue), and the groundtruth (black). Plots (c \& d) show zooms for the most difficult region. The old Gaussian-based method gives wildly inaccurate estimates ($t/\text{Px}=400$) where the BPD profiles are occluded and so measure a non-Gaussian profile. The RRF yields much more accurate estimates, and is improved further (relative to the groundtruth) when fused with the AR (see d).}}
  \label{fig:easyWobble}
\end{figure}

For the ``easy'' scenario (Fig.~\ref{fig:easyWobble}), the RRF instantaneous estimate (green) is mostly accurate, with most estimates close to the groundtruth (black). The old Gaussian-based method (red) gives wildly inaccurate estimates when the BPD is occluded by an object thus resulting in a non-Gaussian BPD profile. However, the RRF yields estimates much closer to the groundtruth, in these cases. These estimates are made very accurate when fused with the AR (blue), since the RRF trees give variable responses which results in a larger uncertainty, so the fusion gives more weight to the AR. In particular, in Fig.~\ref{fig:easyWobble}(d) the fused estimate is much closer to the groundtruth than the RRF on its own.

\begin{figure}[htbp]
   \centering
\includegraphics[width=0.9\textwidth]{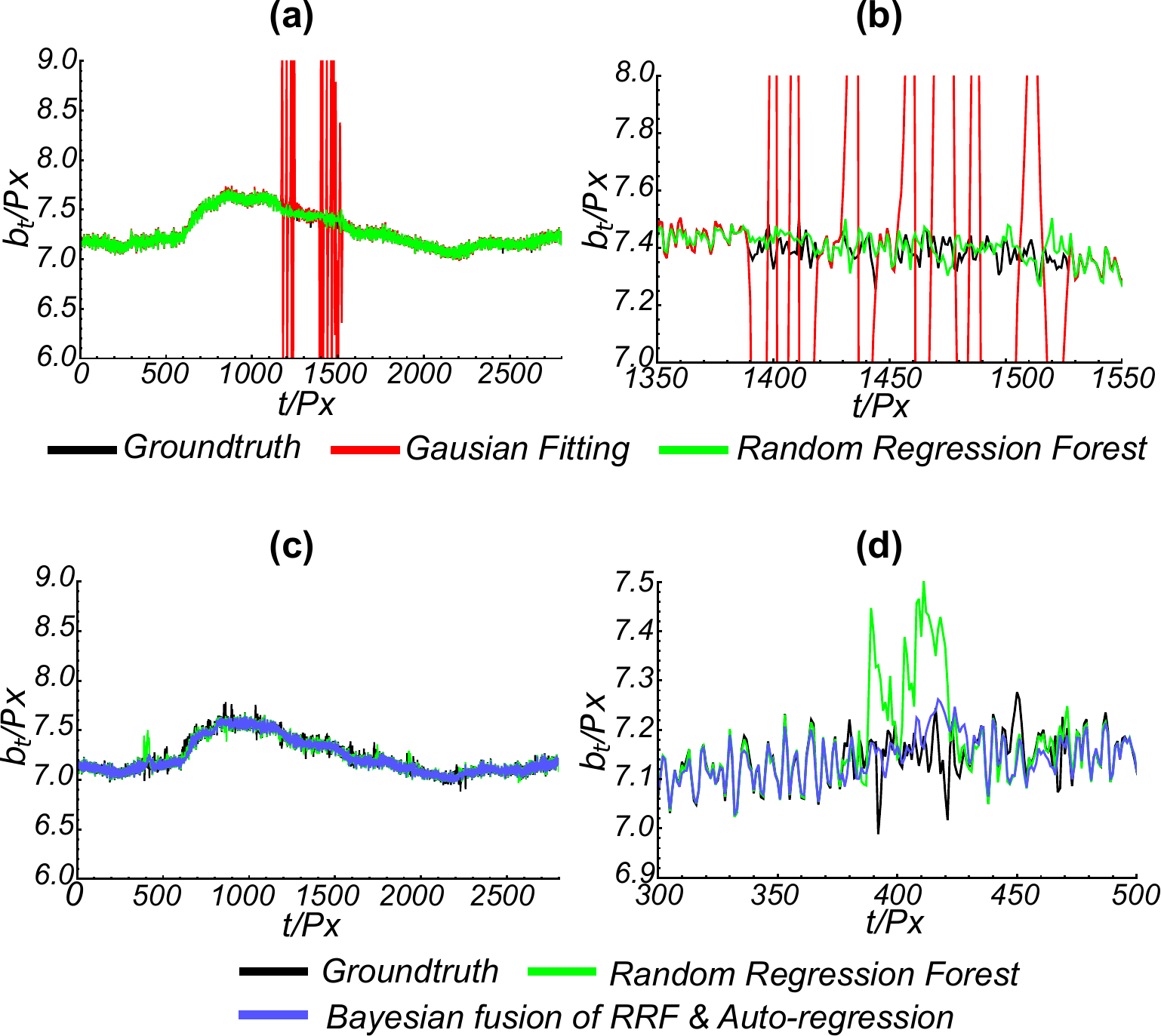}
     \caption{{Beam position estimates for the ``intermediate'' scenario. In (a) the new Random Regression Forest (RRF) based method (green) for instantaneous estimation is compared to the old Gaussian-based method (red), and the groundtruth (black). In (c) the RRF-based method (green) for instantaneous estimation is compared to its Bayesian fusion with an Auto-Regression (AR; blue), and the groundtruth (black). Plots (b \& d) show zooms for the most difficult regions. In this case, the old Gaussian-based method behaves very erratically ($t/\text{Px}=[1100,1600]$), the RRF method gives estimates much closer to the groundtruth. However, some RRF-estimates are inaccurate (d), but these are improved when fused with the AR since the RRF trees give a larger uncertainty than the AR prior estimate.}}
  \label{fig:intermediateWobble}
\end{figure}

In the ``intermediate'' scenario (Fig.~\ref{fig:intermediateWobble}). The old Gaussian-based method does even worse, and again the RRF-based method appears relatively robust to non-Gaussian BPD profiles, where the Gaussian-based method fails. In this scenario, fusion with AR, does not give a large change in estimates over just using the RRF since the RRF trees are confident in their estimate; there is not a large amount of variability in their votes. However, an improvement is seen in Fig.~\ref{fig:intermediateWobble}(d).

\begin{figure}[htbp]
   \centering
\includegraphics[width=0.9\textwidth]{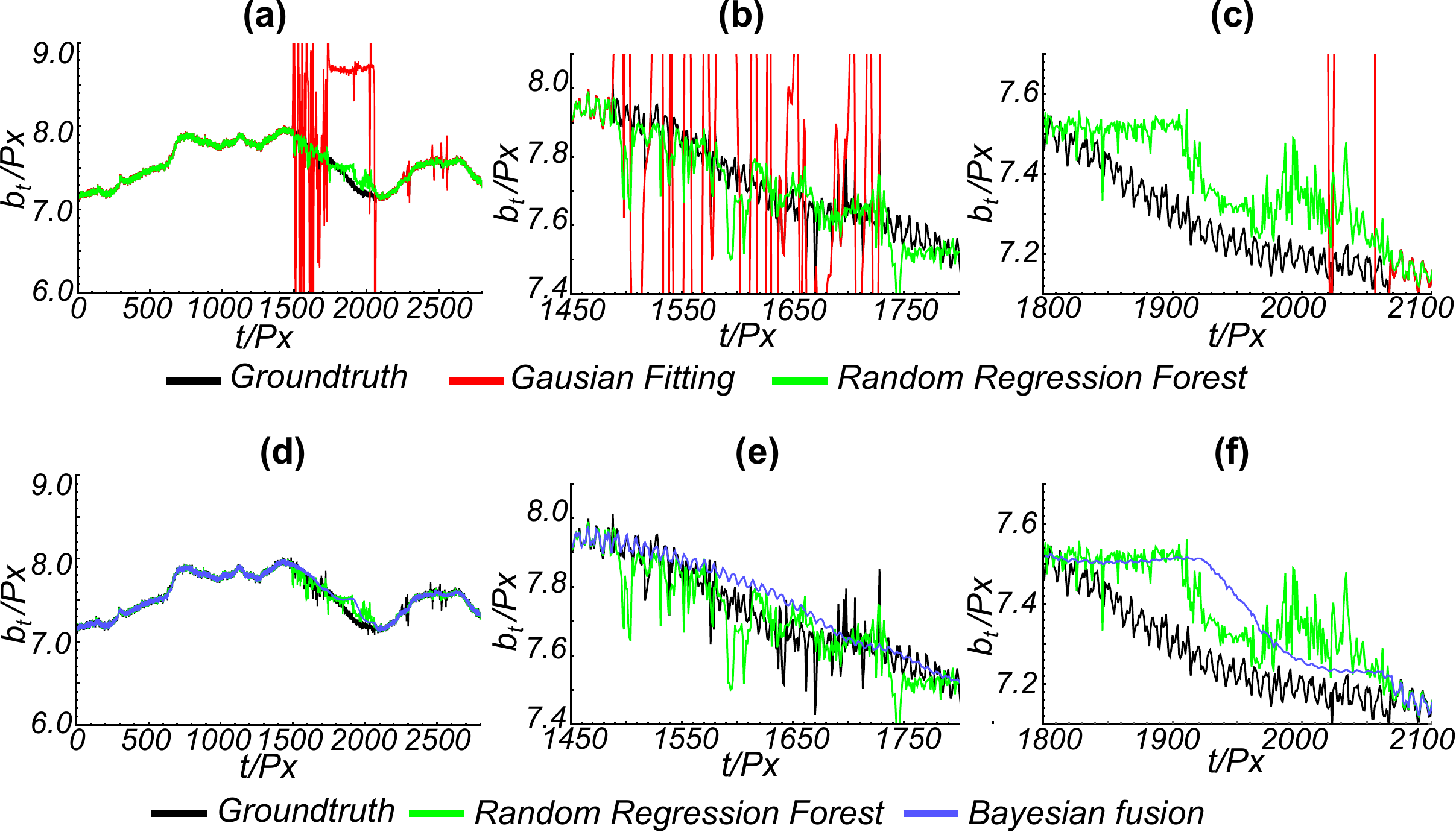}
     \caption{{Beam position estimates for the ``difficult'' scenario. In (a) the new Random Regression Forest (RRF) based method (green) for instantaneous estimation is compared to the old Gaussian-based method (red), and the groundtruth (black). In (d) the RRF-based method (green) for instantaneous estimation is compared to its Bayesian fusion with an Auto-Regression (AR; blue), and the groundtruth (black). Plots (b, c, e, \& f) show zooms for the most difficult regions. The RRF struggles to give accurate estimates in (c), because the BPD is passing across a truck engine, which is very dense and therefore the signal-to-noise is very low. This increases the RRF uncertainty, and so the fused estimate puts full weight on the AR estimate, which results in a constant estimate (f) until a better RRF estimates are achieved. So the AR has forced the Bayesian fusion into giving sensible estimates. This also occurs in (b) and (f) but to a much lesser extent.}}
  \label{fig:difficultWobble}
\end{figure}

For the ``difficult'' scenario in Fig.~\ref{fig:difficultWobble}, the old Gaussian-based method does even worse. The RRF-based instantaneous estimator appears far more robust, with estimates much closer to the groundtruth, however the performance is not as great as in the ``easy'' and ``intermediate'' scenarios. The fused estimates exhibit a bias (see \ref{fig:difficultWobble}.f) where the RRF performs poorly over a long time period. This happens where the total signal on the BPD is close to the background noise level (it is very heavily occluded by a truck engine), and hence the RRF finds it difficult to make accurate estimates of the beam position. This is reflected in the RRF uncertainty, and so the fused estimate puts full weight on the AR estimate, which results in a constant fused beam position estimate until a good instantaneous estimate is achieved. So the AR has forced the fused estimate into giving sensible estimates. Since the BPD signal is so low in this object and it occupies a large number of time-points, we reason that it would be impossible to obtain an accurate instantaneous estimate by any method based on the current BPD set-up.

{For each of the scenarios (easy, intermediate, and difficult), we have quantified the performance of the methods in terms of: accuracy; bias; precision; and Mean Absolute Error (MAE) for the worst 5\%, 1\%, and 0.1\% of time-points. We include the the worst MAEs since particularly bad time-points can be lost in the accuracy, precision, and bias metrics, particularly if there are many air-only time-points where estimation is straightforward. Moreover, wildly inaccurate wobble estimates could lead to column artefacts in the image after correction so are undesirable. The results are given in Table~\ref{tab:wobblePerformance}. For the ``intermediate'' and ``difficult'' scenarios, the RRF-based instantaneous estimation offers roughly an order-of-magnitude improvement across all metrics, over the old Gaussian-based method. For the ``easy'' scenario, this improvement is approximately 3-fold; the Gaussian method is already quite good at dealing with simple objects. By fusing the RRF with the AR (RRF-AR), the performance increases across most metrics, particularly for worst MAEs, however, there is little change (or a slight worsening for the ``intermediate'' scenario) in the overall accuracy. In the ``easy'' scenario there is roughly a 15\% improvement in the MAE for the worst 5\% of time-points. For the ``intermediate'' case the improvement drops so about 5\%. Finally, for the ``difficult'' scenario the worst 1\% MAE improves by about 3\%.}

\begin{table}[!h]
\small
\centering
\caption{Performance metrics for: (i) the old Gaussian (Gauss.) based method of instantaneous estimation; (ii) the proposed Random Regression Forest based method for instantantaneous estimation (RRF); and (iii) the Bayesian fusion of the RRF estimates with an Auto-Regression (RFF-AR). The metrics computed are: Accuracy (Acc.); Bias; Precision (Prec.); and the Mean Absolute Error (MAE) for the worst 5\%,1\% and 0.1\% of estimates. RRF gives an order-of-magnitude improvement over Gauss. for easy and intermediate scenarios, and 3-fold for the difficult scenario. RRF-AR gives 3-15\% improvement in MAE depending on the difficulty.}
\begin{tabular}{llcccccc}
\hline
\bf{Scenario}&\bf{Meth.}&\bf{Acc.} & \bf{Bias} & \bf{Prec.} & \bf{5\%} & \bf{1\%} & \bf{0.1\%}\\\hline

&Gauss.&0.105 & \bf{0.003} & 0.105 & 0.300 & 0.818& 2.085\\
Easy &RRF&0.033 & 0.004 & 0.033 & 0.122 & 0.168& 0.203\\
&RRF-AR&\bf{0.030} & 0.006 & \bf{0.029}& \bf{0.104} & \bf{0.140} & \bf{0.167} \\\hline

&Gauss.& 0.470& -0.010 & 0.470 & 1.751&3.523&5.846\\
Intermediate &RRF&\bf{0.019} & -0.001 & \bf{0.019} & 0.073 & 0.125 & 0.170\\
&RRF-AR&0.021 & \bf{-0.001} & 0.021 & \bf{0.069} & \bf{0.111} & \bf{0.149}\\\hline

&Gauss.&0.637 & -0.135 & 0.623 & 2.012&3.329&5.670\\
Difficult &RRF&0.052 & \bf{-0.008} & -
0.052 & \bf{0.188} & 0.262 & 0.310\\
&RRF-AR&\bf{0.052} & -0.014 & \bf{0.050} & 0.191 & \bf{0.253} & \bf{0.284}\\\hline

 \hline
\end{tabular}
\label{tab:wobblePerformance}
\end{table}

\subsection{Image correction}
{We first assess the image correction method on an air-only scene. For air-only images, wobble estimation is straightforward, since the BPD profile is not distorted by obscuring objects in the scene. However, air-only images allow us to visualise and fully quantify the improvement from wobble correction.} We can assess image quality based on the fact that a perfect (normalised) transmission air image would have all pixel values equalling unity. Image precision can therefore be assessed by computing the {R}oot-{M}ean-{S}quare (RMS) deviation or {P}eak {S}ignal-to-{N}oise {R}atio (PSNR) from this ideal.

\begin{figure}[htbp]
   \centering
\includegraphics[width=0.9\textwidth]{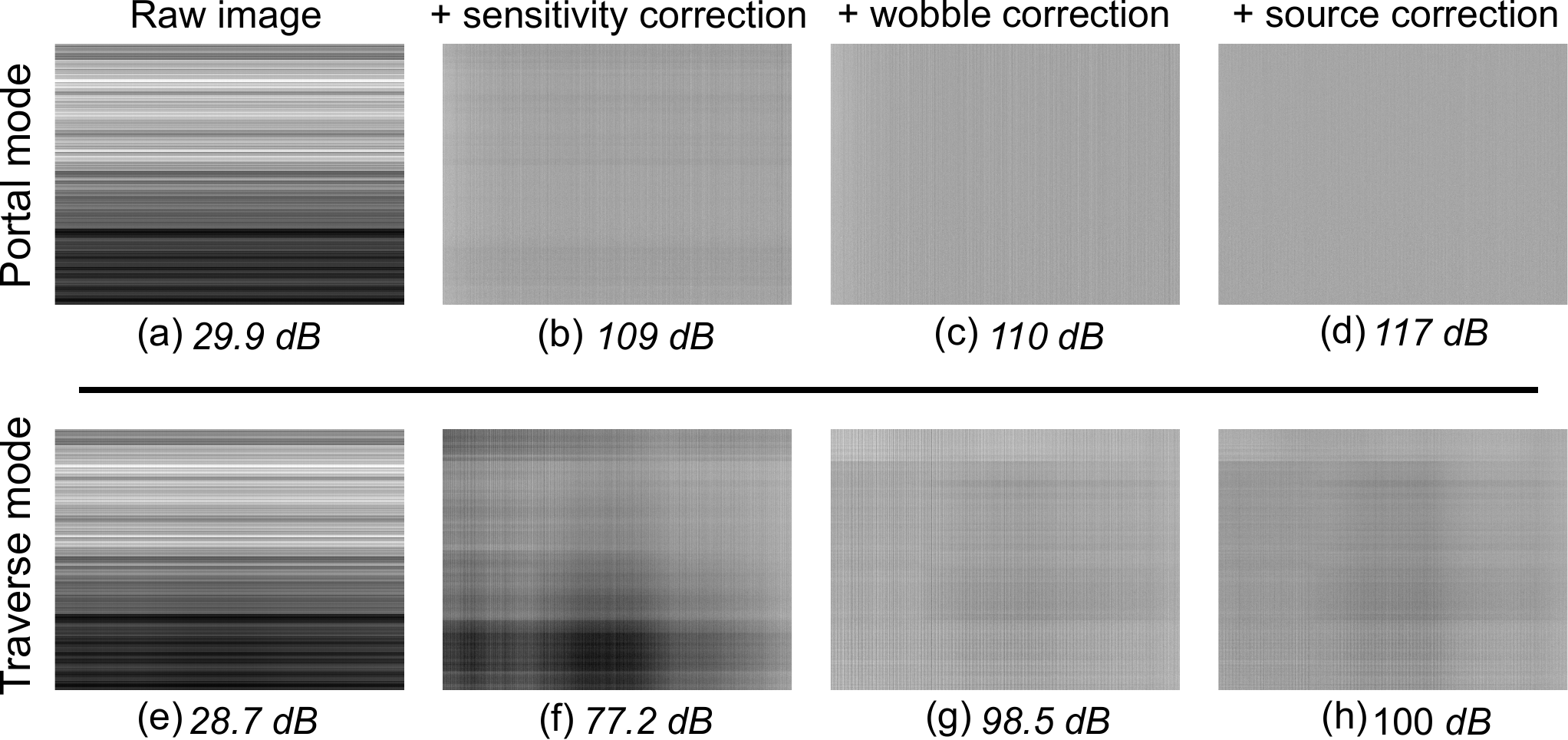}
     \caption{{Images of portal (a-d) and traverse (e-h) mode air-only scans at different stages of correction, including: the raw image; the image after correction for sensor sensitivities; the image after correction for wobble (and sensor offsets); and the final image after after source correction. Corrected images  have been intensity windowed so that the wobble artefact is visible in (b). For each image, the Peak Signal-to-Noise Ratio (PSNR) is given in decibels (dB). A noiseless and artefact-free air-only image should be uniform. The wobble artefact is clearly visible in the traverse image after the sensitivities have been corrected (f), and it is not visible in the portal mode image (b) since this mode is not effected by wobble. The PSNR is reduced by $31.8\,\text{dB}$ by the wobble artefact. After wobble correction (g) there is a visible improvement in the artefact, and the PSNR has improved by $21.3\,\text{dB}$.}}
  \label{fig:airCorrections}
\end{figure}

In Fig.~\ref{fig:airCorrections}, we show air-only images from traverse and portal mode scans and their full correction split into stages. The stages are: sensitivity $R_y$ correction (Fig.~\ref{fig:airCorrections}.b\&f); wobble and ID offset $\exp{(-(b_{ty}-d_y^2)/2\beta_y^2)}$ correction (Fig.~\ref{fig:airCorrections}.c\&g); and source variation $A_t$ correction (Fig.~\ref{fig:airCorrections}.d\&h). Images have been intensity windowed so that the wobble effect is visible in (Fig.~\ref{fig:airCorrections}.f). Note the visible difference between images (Fig.~\ref{fig:airCorrections}.b) and (Fig.~\ref{fig:airCorrections}.f), this difference is mostly due to wobble. The PSNR drops from $109\,\text{dB}$ to $77.2\,\text{dB}$, from portal image (Fig.~\ref{fig:airCorrections}.b) to traverse image (Fig.~\ref{fig:airCorrections}.f) due to the wobble artefact. After wobble correction, to obtain image (Fig.~\ref{fig:airCorrections}.g), most of the wobble artefact is visibly improved. Indeed, the wobble correction improves the PSNR by $21.3\,\text{dB}$ but is unable to achieve the portal mode PSNR.

\begin{figure}[htbp]
   \centering
\includegraphics[width=0.9\textwidth]{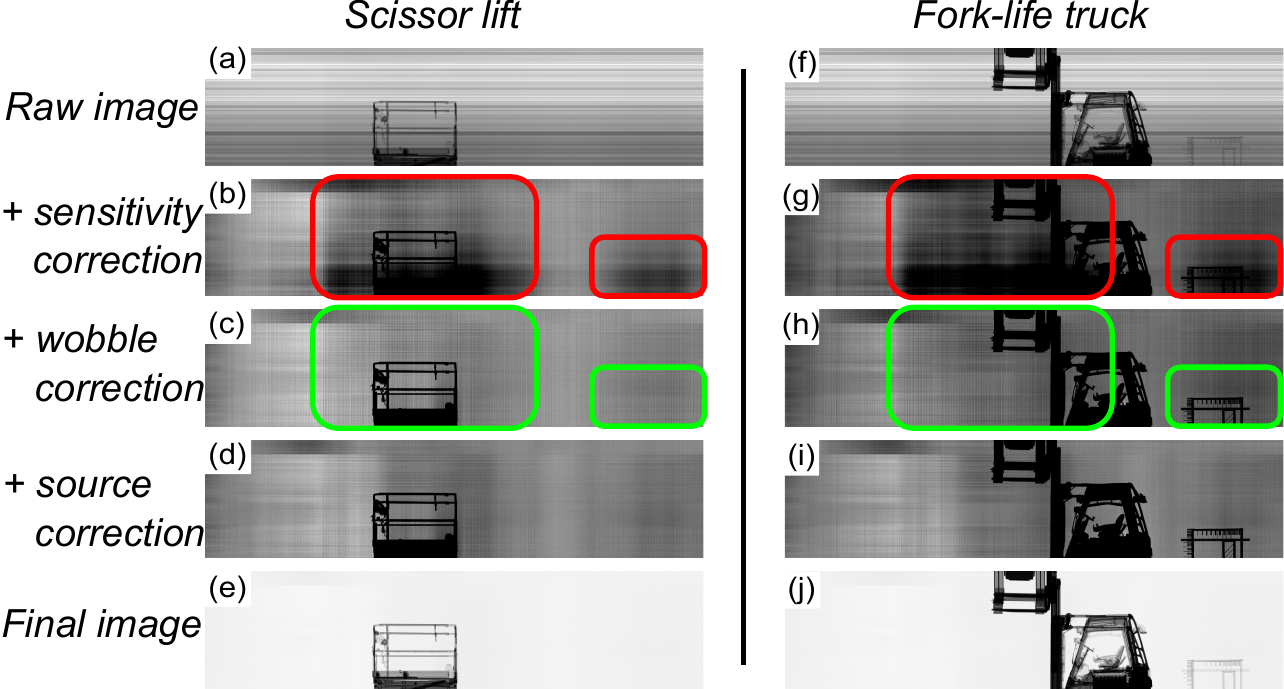}
     \caption{{Images from traverse mode scans of a scissor lift (a-e) and a fork-left truck (f-j), after a series of corrections for: sensor sensitivities (b \& g); wobble and Imaging Detector (ID) offsets (c \& h); and source variation (d \& i). Corrected images  have been intensity windowed so that the wobble artefact is visible in (b) and (g). The final images (e \& j) are the non-windowed versions of (d \& i). The red boxes indicate regions where wobble is particularly visible, and the green boxes indicates the same regions after wobble correction. There is a clear visible improvement in the wobble artefact after correction, and so wobble measurement and correction works quite well even when BPDs are heavily occluded by dense object.}}
  \label{fig:traverseCorrections}
\end{figure}

To make quantitative assessment of the effects visible in Fig.~\ref{fig:traverseCorrections}, the RMS deviations of the traverse and portal mode air-only images, before and after the different corrections, were used to deduce the magnitude of the noise sources before and after correction. Table~\ref{tab:RMSDeviations} shows that wobble increases overall image noise, and has also reduced our ability to correct for \emph{sensor sensitivity}, \emph{ID offset}, and \emph{source} variation. Although it is possible to correct for 99\% of \emph{sensor sensitivity}, the magnitude of \emph{sensor sensitivity} is so large that it is still the second most dominant source of noise in the corrected image. Source variation was the least successfully corrected and this is apparent in Fig.~\ref{fig:traverseCorrections}, since the corrected images (Fig.~\ref{fig:airCorrections}.d\&h) have some slightly visible column artefacts. Finally, we are able to correct $87\%$ of wobble, thus outperforming our previous work~\cite{Rogers2014}, which did not incorporate sensor offset estimates into the correction.

\begin{table}[!h]
\small
\centering
\caption{RMS deviation contributions from different noise sources before and after corrections for: sensor sensitivity; Imaging Detector (ID) offsets; wobble; and source fluctuation. We do not attempt to correct Poisson noise in the photon counts.}
\begin{tabular}{l|l|c|ccc}
\hline
\bf{Scan mode} & \bf{Noise source} & \bf{Symbol} & \bf{Before} & \bf{After} & \bf{Reduction}\\\hline
& \emph{sensor sensitivity} & $R_y$ & 0.2305 & 0.0000 &  100\% \\
& \emph{offset of ID endpoints} & $\{\delta_l,\delta_u\}$ & 0.0013 & 0.0000 & 100\% \\
portal & \emph{wobble} & $b_{ty}$ & 0.0000 & 0.0000 & -- \\
& \emph{source variation}& $A_{t}$ & 0.0030 & 0.0000 & 100\% \\
& \emph{photon count} &-- & 0.0029 & 0.0029 & 0\% \\ \hline
& \emph{sensor sensitivity} & $R_y$ & 0.2305 & 0.0026 & 99\%\\
& \emph{offset of ID endpoints} & $\{\delta_l,\delta_u\}$ & 0.0013 & 0.0004 & 72\%\\
traverse & \emph{wobble} & $b_{ty}$ & 0.0185 & 0.0054 & 87\%\\
& \emph{source variation}& $A_{t}$ & 0.0030 & 0.0004 & 74\%\\
& \emph{photon count}&-- & 0.0029 & 0.0029 & 0\% \\ \hline \hline
\end{tabular}
\label{tab:RMSDeviations}
\end{table} 

The results for corrections applied to traverse mode images of a scissor lift and a forklift truck are shown in Fig.~\ref{fig:traverseCorrections}. Images have been intensity windowed, to the same range, to make the wobble artefact visible. The wobble correction is obtained using the Bayesian-fused beam position estimate. The red boxes indicate image regions most effected by wobble, and the green boxes show the same regions but after wobble correction. There is a visible improvement in the wobble artefact after wobble correction, showing that a good level of correction is obtained even when the BPDs pass through dense objects such as a fork-lift truck.

\begin{figure}[htbp]
   \centering
\includegraphics[width=0.9\textwidth]{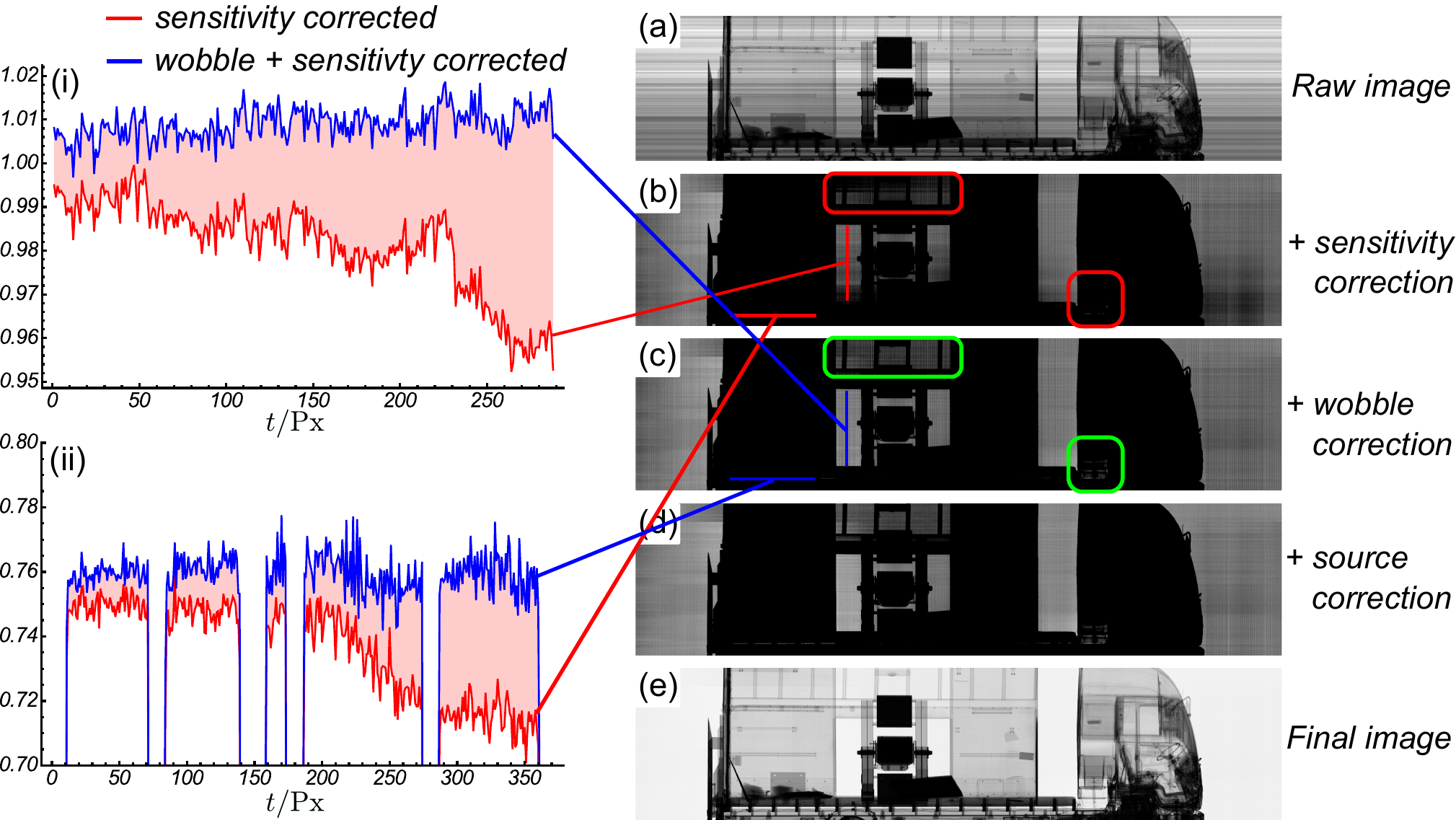}
     \caption{{Images from a traverse mode scan of a truck, after a series of corrections for: sensor sensitivities (b); wobble and Imaging Detector (ID) offsets (c); and source variation (d). Corrected images have been intensity transformed so that the wobble effect is visible in (b).The final image (e) is the non-windowed version of (d). The red boxes indicate regions around the driver's steps and a test object, where wobble is particularly visible. The green boxes indicates the same regions after wobble correction and one can see a visible improvement in the wobble artefact. The plots (i) and (ii) show plots of the pixel intensities across a column and row in the image, respectively. The red traces are uncorrected for wobble and taken from (b), whilst the blue traces are corrected for wobble and taken (c). The red traces should be approximately (piece-wise in ii) constant, however they are distorted by wobble. The wobble correction corrects most of this distortion.}}
  \label{fig:truckCorrections}
\end{figure}

Fig.~\ref{fig:truckCorrections} shows image corrections on a truck image. Since the truck occupies most of the image, it is more difficult to see the effects of wobble and the corrections. The most obvious places are the steps up to the driver's cabin and the area surrounding the test object. {These are indicated by the red boxes in Fig.~\ref{fig:truckCorrections}.b. After wobble correction (green boxes in Fig.~\ref{fig:truckCorrections}.c), the artefact is reduced so that the driver's steps and the test object become visible. In plots Fig.~\ref{fig:truckCorrections}.i\&ii we plot a column and row of pixels, respectively. In each, the red plot is from Fig.~\ref{fig:truckCorrections}.b before wobble correction, and the blue plot is from Fig.~\ref{fig:truckCorrections}.c after wobble correction. The pixels are taken from image lines that should have approximately constant (or piece-wise constant) pixel values. However, due to the wobble artefacts they are distorted from constancy. The wobble correction corrects a large part of this distortion.}

\section{Conclusion}
{We have proposed a series of image corrections to ameliorate detector wobble artefacts in large-scale transmission radiography. The corrections were derived by considering a model of X-ray image formation in the presence of a wobbling detector. The correction relies on the estimation of a number of fixed system parameters and dynamic parameters which vary during a scan. The fixed parameters include sensor sensitivities, sensor misalignments, and the width of the X-ray fan-beam. The dynamic parameters include the position of the beam at different points along the detector array, and the fluctuation of the number of photons emitted by the source. We proposed a method for estimating the fixed system parameters by model fitting to an air calibration image.}

{Wobble is more difficult to estimate, and we adopt a similar approach, using Beam Position Detectors (BPDs), to our previous work~\cite{Rogers2014}. BPDs are placed perpendicular to the imaging array, and measure the cross-sectional profile of the photon beam after interaction with the scene, allowing the position of the beam to be determined and hence detector wobble to be measured. In our previous work, we measured wobble by fitting a Gaussian model to the beam profile to extract an instantaneous estimate of the beam position. This was Bayesian-fused with a prior estimate based on an Auto-Regression (AR). In this contribution, we proposed a new instantaneous estimator based on a Random Regression Forest (RRF). We first estimate the true beam profile, as if the beam had not been attenuated by the scene, and then estimate the beam position and its uncertainty by taking the mean and standard deviation of the responses from a RRF, respectively.}

{To test the wobble estimation and image correction methods, we collected image data of several objects ranging in difficulty from a small scissor lift to a large truck. We used a commercial scanner, which we modified by rotating four imaging detectors by $90^\circ$ to act as BPDs.} Our new RRF-based approach to instantaneous estimation performs significantly {(an order of magnitude in most cases)} better than Gaussian fitting~\cite{Rogers2014}. Moreover, its fusion with an AR achieves results close to ground truth, even for difficult objects{, and performs better than the RRF by 3-15\% in the worst cases.} It struggles for cases where the object has a low signal-to-noise ratio for long durations in the scan, and we believe that this problem cannot be solved {by wobble estimation based solely on BPD readings, unless one can accurately predict future beam positions from a limited number of accurate prior position estimates. This is unlikely due to the stochastic nature of wobble originating from uneven scanning surfaces or wind. Incorporation of measurement devices, such as accelerometers placed along the imaging array, may improve estimates even where there is almost no BPD signal due to object occlusion. This will be a focus of future work.}

The wobble and system parameter estimates were used to apply corrections to images. We applied corrections to traverse and portal mode air-only images and achieved a reduction of $87\%$ of image error due to detector wobble, thus improving on our previous work~\cite{Rogers2014}. The wobble correction method was also applied to difficult images of objects and a {notable qualitative improvement} in the intensity-windowed image quality was observed{, clarifying dense regions of the scene and mitigating human error}. {The method should also allow for improved material discrimination in images captured from dual-energy scanners in traverse mode. State-of-the-art material discrimination, for cargo, is performed by taking the log-ratio (or difference) of images at different energies, and relies on subtle differences between the images~\cite{Ogorodnikov2002,zhang2005hl,ogorodnikov2013material}. But in commercial traverse-mode systems material discrimination is often inaccurate due to image noise, including from wobble (Fig.~\ref{fig:wobble}). And so wobble correction as pre-processing step could help improve material discrimination accuracy. Testing this, and fully quantifying the effect of wobble on material discrimination, will be left to future work.}

Future work, will include experimenting with other measurement devices, such as accelerometers, to improve the prior estimate of the beam position in cases where the RRF fails to obtain accurate estimates over long time-periods due to large, dense objects such as a truck engine. Potentially, beam position estimation could be improved by using more BPDs or even a 2D imaging array, and this will be explored. Additionally, we will investigate the severity and correction of geometric distortions cause by extremely heavy wobble. Such distortions can cause straight lines to become wobble, which potentially impacts on the performance of human operators searching for threats, particularly if their shape is distorted in an unnatural way.

\section*{Acknowledgements}
Funding for this work was provided through the EPSRC
Grant no. EP/G037264/1 as part of UCL's Security Science Doctoral Training Centre, and Rapiscan Systems Ltd.

\bibliography{references}

\end{document}